\documentclass[runningheads]{llncs}

\usepackage[T1]{fontenc}
\usepackage{graphicx}
\usepackage{xcolor}
\usepackage[table, dvipsnames]{xcolor}
\usepackage{newfloat}
\usepackage{listings}
\usepackage{array}

\newcolumntype{C}[1]{>{\centering\arraybackslash}p{#1}}

\definecolor{codegreen}{rgb}{0,0.6,0}
\definecolor{codegray}{rgb}{0.5,0.5,0.5}
\definecolor{codepurple}{rgb}{0.58,0,0.82}
\definecolor{backcolour}{rgb}{0.95,0.95,0.92}

\lstdefinestyle{mystyle}{
    backgroundcolor=\color{backcolour},   
    commentstyle=\color{codegreen},
    stringstyle=\color{codepurple},
    basicstyle=\ttfamily\scriptsize,
    breakatwhitespace=false,         
    breaklines=true,                 
    captionpos=b,                    
    keepspaces=true,                 
    numbers=left,                    
    numbersep=5pt,                  
    showspaces=false,                
    showstringspaces=false,
    showtabs=false,                  
    tabsize=1,
    xleftmargin=2em,
    aboveskip=3pt,
    belowskip=3pt
}
\usepackage{cite}
\usepackage{multirow}
\usepackage[table]{xcolor}
\usepackage{paralist}
\usepackage{arydshln}
\usepackage{enumitem}
\usepackage[table, dvipsnames]{xcolor}

\usepackage{caption}
\usepackage{pifont}
\usepackage{graphicx}
\usepackage{subcaption}
\begin{document}

\title{Revisiting the Role of Natural Language Code Comments in Code Translation}
\titlerunning{Role of Comments in Code Translation}

\author{Monika Gupta\inst{1} \and
Ajay Meena\inst{1} \and
Anamitra Roy Choudhury \inst{2} \and
Vijay Arya \inst{2} \and
Srikanta Bedathur \inst{1}}

\authorrunning{M. Gupta et al.}
%
\institute{Indian Institute of Technology, New Delhi, India \and
IBM Research - India, India}

\maketitle

\begin{abstract}
The advent of large language models (LLMs) has ushered in a new era in automated code translation across programming languages.
Since most code-specific LLMs are pretrained on well-commented code from large repositories like GitHub, it is reasonable to hypothesize that natural language code comments could aid in improving translation quality.
 Despite their potential relevance, comments are largely absent from existing code translation benchmarks, rendering their impact on translation quality inadequately characterized.
 In this paper, we present a large-scale empirical study evaluating the impact of  comments on translation performance. Our analysis involves more than $80,000$ translations, with and without comments, of $1100+$ code samples from two distinct benchmarks covering pairwise translations between five different programming languages: C, C++, Go, Java, and Python. Our results provide  strong evidence that code comments, particularly those that 
 describe the overall purpose of the code rather than line-by-line functionality, significantly enhance translation accuracy. 
Based on these findings, we propose \textit{COMMENTRA}, a code translation approach, and demonstrate that it can potentially \emph{double} the performance of LLM-based code translation. To the best of our knowledge, our study is the first in terms of its comprehensiveness, scale, and language coverage on how to improve code translation accuracy using code comments.
\end{abstract}
\section{Introduction}
\vspace{-6pt}
Accurate and effective source code translation between programming languages (PLs) is a significant challenge for AI. 
Large language models (LLMs) such as CodeLlama \cite{codellama}, DeepSeekCoder \cite{deepseekcoder}, StarCoder \cite{starcoder},  etc., are models pre-trained on a large set of code samples in multiple PLs and offer a potentially faster, less expensive, and more scalable way to automate code translation.
The long-term aim of automated code translation is to handle repository-scale codebases of unbounded complexity; however, we are far from reaching that goal, as LLMs still struggle to consistently generate accurate and reliable translations, even for self-contained programs. 

Several active, ongoing research works (Section~\ref{sec:related}) focus on how to reduce translation bugs by providing relevant context that helps LLMs understand and execute the translation task better.
Since LLMs are typically pre-trained using code and its associated pairwise comments \cite{roziere2020unsupervised}, we suspect that LLMs may translate better when the original source code to be translated is augmented with corresponding natural language code descriptions or code comments. In this work, we first initiate a study of the role that natural language code comments play in LLM-based code translation. 
We address the following research questions:
\begin{itemize}[noitemsep,topsep=2pt]
\item \textbf{\textit{RQ1 - Usefulness of Code Comments:}}
Do natural language code comments help LLMs improve their code translation performance? 
\item \textbf{\textit{RQ2 - Intent of Code Comments:}}
Comment intent refers to the underlying purpose, motivation, or goal a user or developer has when making a comment.
Can we classify code comments based on their intent and comprehend the utility of different intent categories in code translation?
\item \textbf{\textit{RQ3 - Density, Language of Code Comments:}}
Comment density is 
the percentage of comment lines relative to total lines of code.
What role does the density of code comments play in code translation? Also, does the choice of comment language, English or others, affect translation accuracy? 
\item \textbf{\textit{RQ4 - Location of Comments:}}
Code comments are typically placed in specific locations to provide context and explanations for human readers, while being ignored by the compiler or interpreter.
Natural language pseudo-code can be placed at the beginning of the code; method specifications are usually placed at the method boundaries; and more fine-grained comments are positioned next to the line(s) of code with which they are associated. How does this placement affect translation performance?
\end{itemize}
To address these research questions, we collected 1100+ code samples in five different PLs, C, C++, Go, Java, Python, augmented them with model-generated comments and translated each commented and uncommented sample into other PLs, using several LLMs, obtaining more than $80K$ code translations. Our analysis indicates that while comments can enhance the code translation performance of LLMs, their indiscriminate usage can have a deleterious effect. 
We next propose a translation framework called \textbf{COMMENTRA}, which carefully introduces comments in programs to significantly improve the performance of LLM-based code translation.
Experiments show that COMMENTRA offers an impressive 
gain in average translation performance across all PLs, benchmark datasets, and LLM models considered in our study. 
The attractiveness of COMMENTRA lies in its simplicity, low-cost implementation, and ease of integration with many recently proposed code translation frameworks, such as \textit{InterTrans} \cite{intertrans}, \textit{LANTERN} \cite{lantern}, \textit{UniTrans} \cite{intertrans}, \textit{UniTranslator} \cite{unitranslator}, etc. 
To the best of our knowledge, our work is the first comprehensive study of code comments in the LLM-based code translation process across a wide configuration.
\section{Related Work}
\label{sec:related}
\vspace{-6pt}
\noindent\emph{\textbf{Code Translation Approaches.}} 
Rule-based, language-specific transpilers use explicitly defined linguistic and syntactic rules to convert source code from one PL to another. While transpilers are reliable for known patterns, they struggle with complexity and generality. In contrast, LLMs are trained on data, produce more idiomatic code, but lack formal guaranties and can fail unpredictably. Examples of transpilers include 
Java2Python\footnote{\url{https://github.com/natural/java2python}}, 
Python2Java\footnote{\url{https://github.com/chrishumphreys/p2j}}, 
C2Rust\footnote{\url{https://github.com/immunant/c2rust}}, 
C2Go\footnote{\url{https://github.com/gotranspile/cxgo}}, 
etc.
Techniques such as lexical statistical machine translation \cite{nguyen2013lexical}, tree-based neural networks \cite{chen2018tree}, deep learning, and unsupervised learning \cite{roziere2020unsupervised, lachaux2021dobf} have also been explored for automatic code translation. 
LLM-based methods such as AlphaTrans \cite{alphatrans}, CoTran \cite{jana}, InterTrans \cite{intertrans}, UniTrans \cite{unitrans}, UniTranslator \cite{unitranslator}, etc., have been proposed to improve the LLM's code translation accuracy. We compare these recent approaches with our work in the 
Appendix.

\noindent\emph{\textbf{Use of Code Comments in Training Models.}}
In earlier works, it was observed that keeping comments in the training dataset facilitates cross-language alignment and yields better results for some language pairs during unsupervised code translation \cite{roziere2020unsupervised, dua}. Our work differs from these in that COMMENTRA does not use code comments to train models. In fact, the hypothesis that LLMs are pre-trained using code and its associated pairwise comments is the motivation behind this work.

\noindent\emph{\textbf{Use of NL Specifications.}} 
Saha et al. \cite{saha} investigated the use of pseudocode, and Tang et al. \cite{tang} explored the usefulness of self-generated natural language explanations as an intermediate representation for code translation. Nitin et al. \cite{spectra} generated multimodal specifications (invariants, tests, descriptions) and used them to augment the source code, thereby improving the translation performance of LLMs. Code comments differ from pseudocode and structured code specifications in their close, precise association with the corresponding line(s) of code. Additionally, while pseudocode and code specifications focus on explaining the basic code logic, code comments, in contrast, can depict various intents \cite{mu}. Our work differs from these as we present the first comprehensive study on the impact of code comments in the code translation process. 
 
\noindent\emph{\textbf{Code Comment Generation.}}
Many models and approaches have been proposed \cite{summarize, gao2, Gao1, gong, codenn} to generate natural language code summaries. A piece of code can be commented on with several intentions \cite{mu2023developerintentdrivencodecomment, zhai}, such as its functionality, its usage, and its space/time complexity, etc.. Mu et al. \cite{mu} propose an intent-driven DOME code comment generation approach to produce a comment that is coherent with the given intent. 
\section{Experimental Setup}
\label{sec:setup}
\begin{figure}[!h]
 \vspace{-10pt}
\centering
\includegraphics[width=0.5\textwidth]
{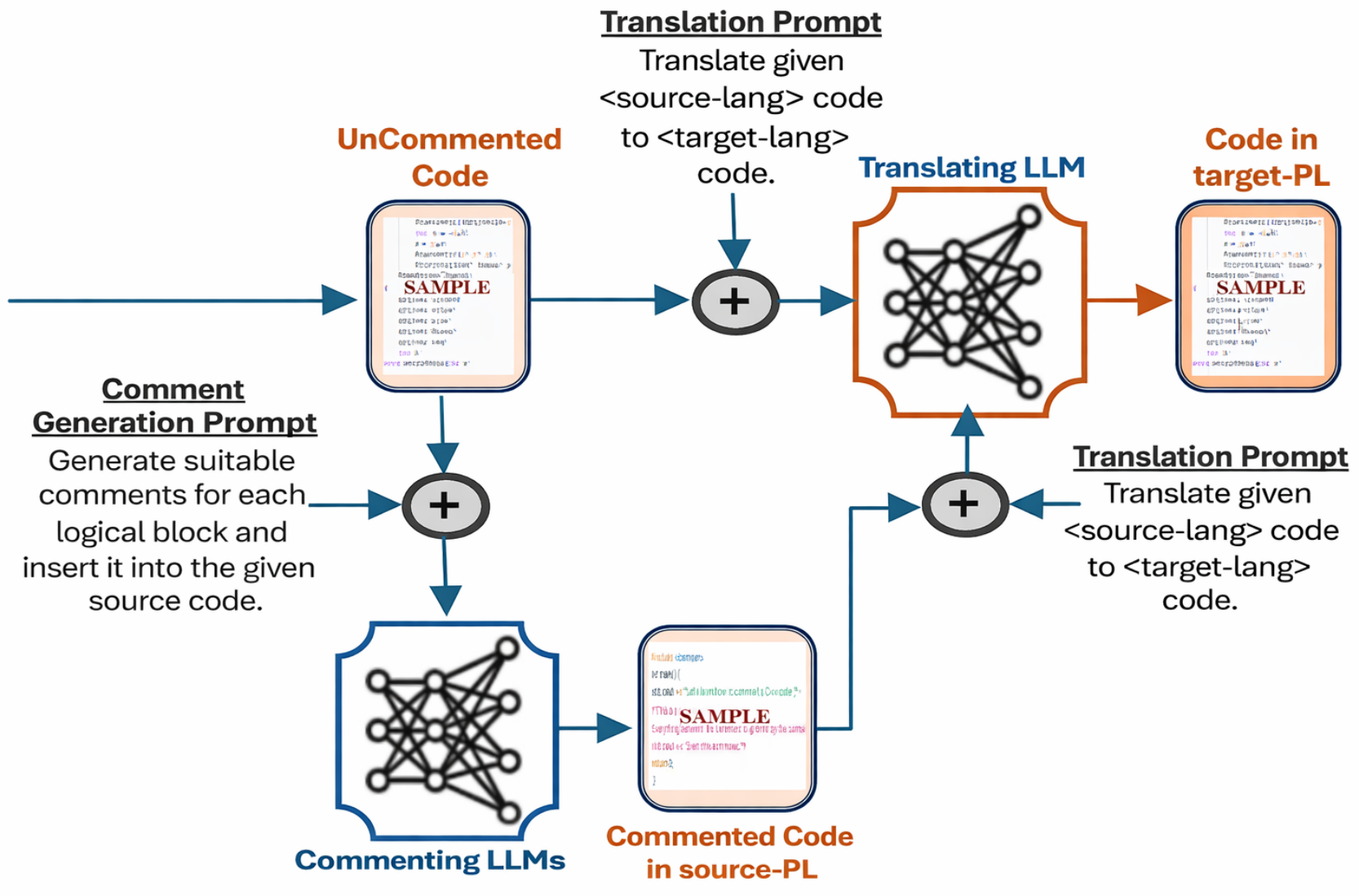}
    \caption{Experimental Setup; The exact prompts used are also shown here.}
    \label{fig:setup} 
    \vspace{-2pt}  
\end{figure}

\vspace{-6pt}
Figure \ref{fig:setup} illustrates our experimental setup. We enrich uncommented code samples with LLM generated comments using a simple comment generation prompt that has no explicit mention of any specific comment intent, density, language, or comment location. Both commented and uncommented code samples are translated using a vanilla translation prompt to generate a large pool of translations.  All prompts were used in their basic \emph{vanilla} form (refer to Figure \ref{fig:setup} for the exact prompts) to minimize potential bias due to the different ways in which various LLMs utilize more sophisticated prompts. A single A100 Nvidia GPU was used for all experiments.

 \vspace{-15pt}
 \textbf{Translating and Commenting LLMs.\hspace{5pt}}
We used \textit{Code Llama-13B-V1} \cite{rozière2024code}, 
\textit{DeepSeek-Coder-V2} \cite{deepseekai2024deepseekcoderv2breakingbarrierclosedsource},
\textit{GPT-4o-mini} \cite{openai},
\textit{Granite-8B-Code-Instruct} \cite{mishra},  and  \textit{StarCoder-1} \cite{li2023starcoder} to translate all code samples. \textit{DeepSeek-Coder-V2}, \textit{GPT-4o-mini}, and \textit{Mistral 7B} \cite{mistral7b}
were used as the Commenting Models. Henceforth, we refer to these models as \textit{CodeLlama, DeepSeek, GPT, Granite, Mistral, and StarCoder}, accordingly. For all models, the token length was set to its maximum, and the greedy decoding algorithm, which selects only the top translation with the highest log probability, was used. 

\begin{table}[!h]
\vspace{-10pt}
\centering
\footnotesize
\begin{tabular}{ccccc}
    \hline
    \multicolumn{5}{c}{\textbf{Dataset Statistics}} \\
    \hline
    Code Set & Source & No. of  & Min-Max LOC & Avg-TestCases\\
    Source  & PL     & Samples & in samples  & per sample\\
    \hline
    AVATAR & Java & 250 & 12 - 78 & 12\\
    AVATAR & Python & 250 & 1 - 64 & 12\\
    CodeNet & C & 200 & 7 - 318 & 1\\
    CodeNet & C++ & 200 & 7 - 194 &  1\\
    CodeNet & Go & 200 & 7 - 393 &  1\\
    \hline
    \multicolumn{5}{l}{Total Unique, Uncommented Code Samples : 1100} \\
    \hline
  \end{tabular}
  \caption{Benchmark Data Statistics} 
\label{tab:benchmarks}
\end{table}
 \vspace{-25pt}
\textbf{Datasets.\hspace{5pt}}
We studied translations of code samples between five different PLs - C, C++, Go, Java, and Python - resulting in a total of 20 (i.e., 5 permute 2) unique \textit{(source-PL, target-PL)} pairs. A total of 1100 unique code samples (see Table \ref{tab:benchmarks}), along with their multiple test cases, from two standard, openly available, widely studied benchmark suites 
- AVATAR \cite{ahmad2021avatar} and CodeNet \cite{puri2codenet} were used for the experiments. 
Since these code samples do not have any  comments in them, we employed
commenting models to generate comments. To rigorously assess the effectiveness of our proposed framework, COMMENTRA, we performed additional experiments on the CodeTransOcean  benchmark \cite{codetransocean}, which is a multilingual reference for evaluating code translations in a variety of programming languages. These additional results are presented in the 
Appendix.

\begin{table}[t]
\centering
\includegraphics[width=\columnwidth]{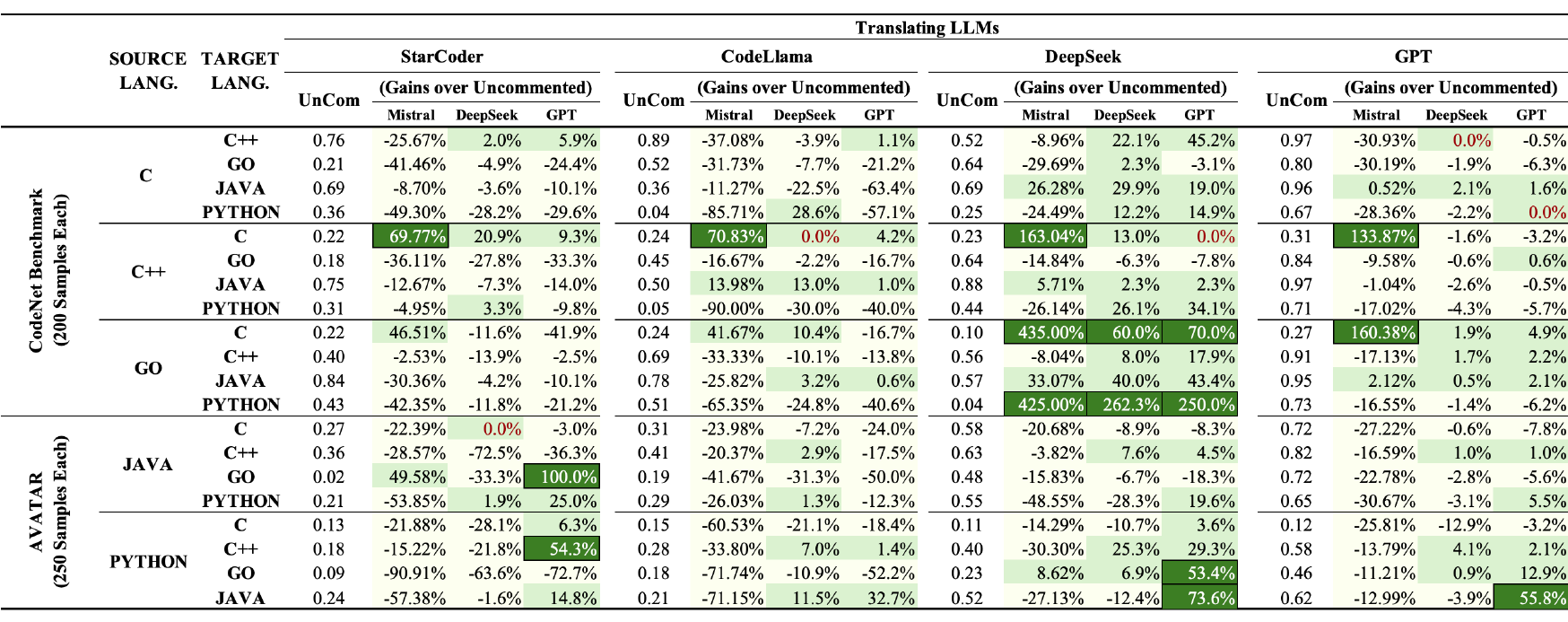}
  \caption{The impact of comments generated by various \textit{commenting LLMs} over baseline translation performance (accuracy) of different \textit{translating LLMs} across different pairwise language translations in benchmark datasets. 
  }
  \label{rq1}
\end{table}
\section{How Relevant are Code Comments?}
\vspace{-6pt}
\subsection{RQ1 : Usefulness of Code Comments}
\label{subsec:rq1}
To explore the usefulness of code comments, we injected all code samples with comments generated by three different commenting LLMs, resulting in 3300 commented code samples (refer to Section \ref{sec:setup} for the various translation and comment generation prompts used in this experiment).
The resultant collection of all uncommented and commented code samples was then translated into other target PLs using five different translating LLMs. 
For each 4-tuple set (\textit{source-PL, target-PL, commenting-LLM, translating-LLM}), the percentage of successful translations was calculated as the ratio of the total number of successful translations achieved to the total number of translations attempted in the set. A translation is successful if the translated code compiles, passes runtime checks, and all existing test cases. For all our code samples, the corresponding input test cases were available as a formatted set of input values (see Listings \ref{lst:ex1-3}, \ref{lst:ex3-3}) as part of the source benchmark they were from, and therefore they could be used as-is for testing both the source program and the translated program. The higher the percentage of successful translations for an LLM, the better the translation performance of the LLM on the corresponding set. Due to space constraints, Table \ref{rq1} presents these results\footnote{Results for all translating LLMs are included in the 
Appendix
.} for a subset of translating LLMs. As an example of how to interpret the numbers in the table, consider the C to C++ translation numbers of StarCoder. The value 0.76 in \textit{UnCom} column is the baseline; it indicates that the translation of 76\% of the total number of code samples (200 in this case) was successful without any comments. With Mistral-comments, there was a 25.67\% drop in StarCoder's translation performance; i.e., the number of successful translations was reduced to 113 out of 200. However, with DeepSeek-comments, the total number of successful translations increased by 2\% from the baseline; that is, 155 of 200 samples were translated successfully. Similarly, with GPT-comments, the number of successful translations rose to 161 out of 200. 

Green colored boxes in the table represent cases where the presence of code comments helped increase the translation performance of LLMs, while the yellow colored boxes indicate the contrary. Additionally, highlighted in dark green are the cases where, after adding code comments, the performance enhancement 
over the baseline was more than 50\%. A nearly balanced distribution of green and yellow boxes is observed. With commented code, the gain in LLM translation performance ranged from -90\% to +435\% over the baseline uncommented code translation performance. This leads us to conclude that while model-generated code comments can help improve the translation performance of LLMs, they can also have a deleterious effect on the translations. This seems to be due to their injecting more noise or constraining the LLMs in translation logic. 
We discuss several examples in support of this conclusion in Section 4.2. \\
\begin{figure}[t]
\centering
  \includegraphics[width=\columnwidth]{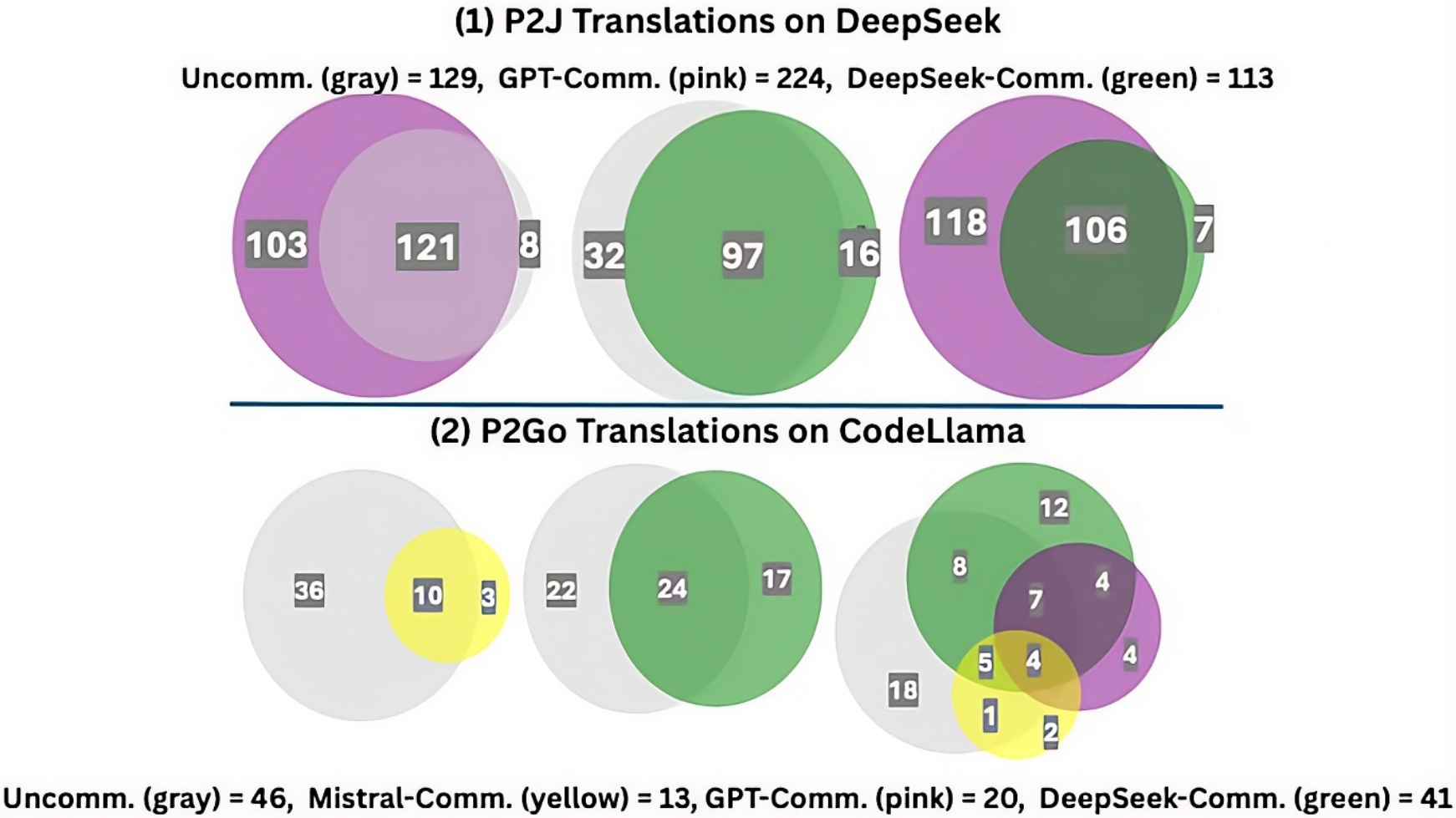}
  \caption{Venn diagrams depicting increase and decrease in LLMs performance in the commented code samples. Left and center diagrams show the overlap between uncommented successful and successfully translated model-commented  samples; the right diagrams show the overlap between the various successfully translated model-commented samples.}
  \label{fig:rq1-inc}
\end{figure}
\noindent\textbf{Understanding the increase in performance cases.}
Figure \ref{fig:rq1-inc}(1) presents Venn diagrams for an in-depth analysis of a representative case where the LLM showed an increase in performance in the commented sets, Python to Java (P2J) translations on DeepSeek. The total number of successful translations in the uncommented set (gray) was 129/250, compared to 224/250 in the GPT-commented (pink) and 113/250 in the DeepSeek-commented (green). The size of the gray bubble is smaller than the size of the pink bubble, indicating that the GPT-commented samples translated better than the uncommented ones; hence, there is an increase in performance in this case. Indeed, it is clear from the Venn diagrams that GPT-comments and DeepSeek-comments added value in at least 103/250 and 16/250 additional code samples, respectively, that were unsuccessful when there were no comments at all. These observations clearly highlight that code comments help to iron out technical flaws that may have occurred while translating uncommented code. Ideally, we would expect the uncommented sets (gray) to be completely contained within the commented sets (other colors). However, the fact that this is not the case indicates that comments can potentially deteriorate translations. \\
\noindent\textbf{Understanding the decrease in performance cases.}
Figure \ref{fig:rq1-inc}(2) presents an in-depth analysis of another representative case - P2Go translations on CodeLlama, where the LLM shows a decrease in performance in the commented sets. The total number of successful translations in the uncommented set was 46/250 (gray), compared to 13/250 in the Mistral-commented (yellow), 20/250 in the GPT-commented (pink), and 41/250 in the DeepSeek-commented (green). Although gray bubbles are larger than colored bubbles, the overlap between them is not very large, indicating the usefulness of code comments in resolving some of the failed uncommented translations. As is clear from the diagram, GPT-comments and DeepSeek-comments added value in at least 8 and 17 additional code samples, respectively. These code samples were previously unsuccessful when there were no comments at all. These observations reaffirm that comments can have both a positive and a negative effect on the translation process. \\
\noindent\textbf{Types of errors resolved / introduced by the presence of code comments.} 
\begin{figure}[t]
\centering
  \includegraphics[width=\textwidth]{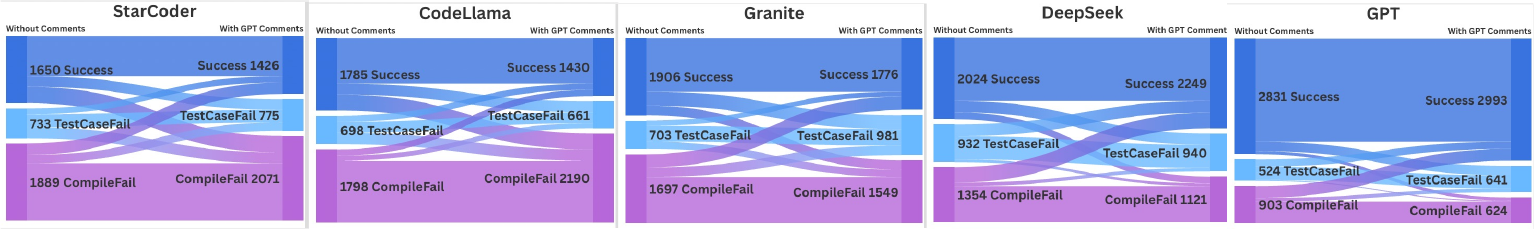}
  \caption{Bird's eye view of how various translation models performed on commented and uncommented code.}
  \label{fig:bug-anal}
\end{figure}
We compared the error-type distributions in GPT-commented translations and uncommented translations for all models (results in Figure \ref{fig:bug-anal}). 
Across all unique \textit{(source-PL, target-PL)} pairs, a total of 4,400 uncommented and 4,400 GPT-commented translations were attempted on each translation model in our study. Each of these translations was either a successful translation or contained an error, such as a testcase failure, compilation failure, or other errors like runtime failures or infinite loop errors, etc. As an example, with the StarCoder model, for uncommented translations, the error type distribution was (Total-Attempted:4400, Successful:1650, Testcase-Fail:733, Compile-Fail:1889, Other errors:128). With GPT-comments, a few samples from the original pool of successful translations began to either fail in testcases or compilation. In addition, a few samples from the original pool of testcase failures and compilation failures were transformed into successful translations.
 Clearly, for all models, across all language pairs, a substantial number of translations that originally (during the uncommented translation run) had compilation errors became successful with GPT-comments. On the contrary, compared to DeepSeek and GPT, in the StarCoder and CodeLlama models, a higher percentage of commented samples, which were originally successful when uncommented, began to fail with compilation errors, possibly hinting at the inability of these models to judiciously understand and use the additional context provided by the comments.\\
  \begin{figure}[t]
\centering
\includegraphics[width=0.5\linewidth]{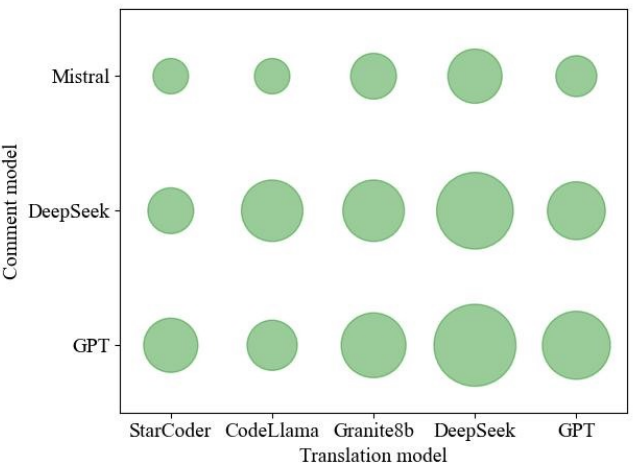}
  \caption{Commenting Model Comparison. Size of the green bubbles $\propto$ no. of \textit{(source-PL, target-PL)} pairs for which the corresponding \textit{(comment-model, translation-model)} pair showed positive gains over the baseline. 
 }
  \label{fig:model-comp}
\end{figure}
\noindent\textbf{Impact of commenting models on translation performance.}
For uncommented code, the translation performance of LLMs depends primarily on the chosen source and target PL. However, in the case of commented code, the specific NL words used in the code comments also influence the translating LLM. Since the performance of commenting models may itself be PL-specific, it is difficult to generalize and choose an overall best translating and commenting model. In our experiments, Mistral as a commenting model showed limited capabilities on many \textit{(source-PL, target-PL)} translation pairs (e.g., Figure \ref{fig:rq1-inc}), although DeepSeek and GPT performed much better (e.g., for P2Go, P2J, P2C++, J2P translations). Figure \ref{fig:model-comp} presents a high-level comparative analysis of the various commenting models by summarizing the results from Table \ref{rq1}, where we count the number of \textit{(source-PL, target-PL)} pairs  that benefited from adding comments. The smaller size of the bubbles in the row corresponding to the Mistral commenting model clearly highlights that DeepSeek and GPT proved to be better commenting models than Mistral for most \textit{(source-PL, target-PL)} pairs. \\
Our key takeaways from this study are:
\begin{itemize}[noitemsep,topsep=0pt]
    \item Although well-written comments that describe the corresponding code are valuable in resolving several translation errors, they can sometimes contribute to noise, leading to translation errors.
    \item The added value of code comments varies across different source and target PLs and also across different translating and commenting LLMs. 
\end{itemize}
\vspace{-10pt}
\subsection{Examples - Positive and Negative Impact of Code Comments}
\vspace{-2pt}
\textbf{Example 1 : Code comments help resolve testcase failure bugs.}
Consider the given P2J translation on Granite, with GPT as the commenting LLM. The uncommented P2J translation compiled, but some test cases failed during execution. However, in the commented P2J translation, all test cases passed.
\begin{lstlisting}[caption=Original Uncommented Python Source Code, label={lst:ex1-1}, language=python]
n = int ( input ( ) )
list = [ int ( x ) % 2 for x in input ( ).split ( ) ]
if sum ( list ) == 1 :
    print ( list.index ( 1 ) + 1 )
else :
    print ( list.index ( 0 ) + 1 )
\end{lstlisting}
The above Python program converts each input integer in the given list into either 0 or 1 based on whether it is odd or even, and stores them in an array. It then sums the array; if it equals 1, it prints the index of the entry that was "1"; otherwise, it prints the index of the first "0" entry. The translation to Java by Granite given below is erroneous in the for-loop in lines 12-17.
\begin{lstlisting}[caption=Output Java Language Translation, label={lst:ex1-2}, language=java]
import java.util.Scanner;
public class Main {
    public static void main(String[] args) {
        Scanner scanner = new Scanner(System.in);
        int n = scanner.nextInt();
        int[] list = new int[n];
        for (int i = 0; i < n; i++) {
            list[i] = scanner.nextInt() % 2;
        }
        int index = 0;
        for (int i = 0; i < n; i++) {
            if (list[i] == 1) {
                index = i;
                break;
            }
        }
        if (index == 0) {
            for (int i = 0; i < n; i++) {
                if (list[i] == 0) {
                    index = i;
                    break;
                }
            }
        }
        System.out.println(index + 1);
    }
}
\end{lstlisting}
\vspace{-5pt}
\begin{lstlisting}[label={lst:ex1-3}, caption=Failed Test Case]
-- Input --
3
1 2 2
-- Expected Output --
1
-- Actual Output --
2
\end{lstlisting}
\vspace{-5pt}
In the commented version of the Python program given below, GPT-comments, especially those in lines 7, 9, and 12, precisely describe the program logic. 
\begin{lstlisting}[caption=GPT-Commented Python Source Code, label={lst:ex1-4}, language=python]
# Read an integer input which represents the number of elements
n = int(input())
# Read a line of input, split it into individual elements, convert them to integers,
# and create a list that contains 1 for odd numbers and 0 for even numbers
list = [int(x) % 2 for x in input().split()]
# Check if there is exactly one odd number in the list
if sum(list) == 1:
    # If there is one odd number, print its position (1-based index)
    print(list.index(1) + 1)
else:
    # else print the position of the first even number (1-based index)
    print(list.index(0) + 1)
\end{lstlisting}
The translating LLM, Granite, is influenced by the GPT-comments and follows the program logic explained by the comments. Lines 11-16 seem to implement the comment in line 7 
; also, the if-else control from line 17 onward seems to result from the comments in lines 9 and 12 of Listing \ref{lst:ex1-4}. As a result, the translation becomes \emph{successful}. We tried the commented P2J translation after removing all code comments except those in lines 7, 9, and 12. The resultant Java output was also a successful translation, which further demonstrates the positive role played by the comments in these lines.
\begin{lstlisting}[caption=Output Java Language Translation, label={lst:ex1-5}, language=java]
import java.util.Scanner;
public class Main {
    public static void main(String[] args) {
        Scanner scanner = new Scanner(System.in);
        int n = scanner.nextInt();
        int[] list = new int[n];
        for (int i = 0; i < n; i++) {
            list[i] = scanner.nextInt() % 2;
        }
        int count = 0;
        for (int i = 0; i < n; i++) {
            if (list[i] == 1) {
                count++;
            }
        }
        if (count == 1) {
            for (int i = 0; i < n; i++) {
                if (list[i] == 1) {
                    System.out.println(i + 1);
                    break;
                }
            }
        } else {
            for (int i = 0; i < n; i++) {
                if (list[i] == 0) {
                    System.out.println(i + 1);
                    break;
                }
            }
        }
    }
}
\end{lstlisting}

\noindent{\textbf{Example 2 : Code comments negatively impact code translation.}}\\
Consider the given J2Cpp Translation on CodeLlama with GPT as the commenting LLM. The uncommented J2Cpp translation was successful; however, in the commented J2Cpp translation, some test cases failed.

\begin{lstlisting}[caption=Original Uncommented Java Source Code, label={lst:ex3-1}, language=C++]
import java.io.BufferedWriter;
import java.io.IOException;
import java.io.OutputStreamWriter;
import java.util.Scanner;
public class Main {
	public static void main(String[] args) throws IOException {
		Scanner sc = new Scanner(System.in);
		BufferedWriter bw = new BufferedWriter(new OutputStreamWriter(System.out));
		int H = sc.nextInt();
		int W = sc.nextInt();
		int A = sc.nextInt();
		int B = sc.nextInt();
		if (W < 2 * A || H < 2 * B) {
			System.out.println(-1);
			return;
		}
		for (int i = 0; i < H; i++) {
			for (int j = 0; j < W; j++) {
				if ((i < B && j < A) || (i >= B) && (j >= A))
					bw.write("0");
				else
					bw.write("1");
			}
			bw.newLine();
		}
		bw.flush();
	}
}
\end{lstlisting}
\vspace{-5pt}
The above Java program inputs four integers: \textit{H}, \textit{W}, \textit{A}, 
and \textit{B}. It checks whether it is possible to construct the matrix; if not, it prints -1 and stops; otherwise, it directly prints the matrix row by row to the output. Its C++ translation (Listing \ref{lst:ex3-2}) captures the intended logic and executes successfully.
\begin{lstlisting}[caption=Output C++ Language Translation, label={lst:ex3-2}, language=go]
#include <iostream>
#include <cstdio>
#include <cstring>
using namespace std;
int main()
{
    int H, W, A, B;
    cin >> H >> W >> A >> B;
    if (W < 2 * A || H < 2 * B)
    {
        cout << -1 << endl;
        return 0;
    }
    for (int i = 0; i < H; i++)
    {
        for (int j = 0; j < W; j++)
        {
            if ((i < B && j < A) || (i >= B) && (j >= A))
                cout << "0";
            else
                cout << "1";
        }
        cout << endl;
    }
    return 0;
}
\end{lstlisting}
\vspace{-5pt}
Consider the commented version (Listing \ref{lst:ex3-4}) of the original Java code, and note the use of the word \textit{grid} in the comments in lines 11, 12, 13, 16, 18, and 21. 
\begin{lstlisting}[caption=GPT-Commented Java Source Code, label={lst:ex3-4}, language=C++]
import java.io.BufferedWriter; 
import java.io.IOException; 
import java.io.OutputStreamWriter; 
import java.util.Scanner; 
public class Main { 
    public static void main(String[] args) throws IOException { 
        // Create a Scanner object for input and a BufferedWriter for output
        Scanner sc = new Scanner(System.in); 
        BufferedWriter bw = new BufferedWriter(new OutputStreamWriter(System.out)); 
        // Read the dimensions of the grid and the parameters A and B
        int H = sc.nextInt(); // Height of the grid
        int W = sc.nextInt(); // Width of the grid
        int A = sc.nextInt(); // Parameter A
        int B = sc.nextInt(); // Parameter B
        // Check if the grid can accommodate the required number of 0s and 1s
        if (W < 2 * A || H < 2 * B) { 
            System.out.println(-1); // Print -1 if the grid is too small
            return; // Exit the program
        } 
        // Generate the grid based on the parameters A and B
        for (int i = 0; i < H; i++) { // Iterate over each row
            for (int j = 0; j < W; j++) { // Iterate over each column
                // Determine whether to write '0' or '1' based on the current position
                if ((i < B && j < A) || (i >= B && j >= A)) 
                    bw.write("0"); // Write '0' for specified conditions
                else 
                    bw.write("1"); // Write '1' otherwise
            } 
            bw.newLine(); // Move to the next line after finishing a row
        }       
        // Flush the BufferedWriter to ensure all data is written out
        bw.flush(); 
    } 
}
\end{lstlisting}
\vspace{-10pt}
\begin{lstlisting}[caption=Output C++ Language Translation, label={lst:ex3-5}, language=go]
#include <iostream> 
#include <cmath> 
using namespace std; 
int main() { 
    // Declare variables for input and output
    int H, W, A, B; 
    char grid[100][100]; 
    // Read the dimensions of the grid and the parameters A and B
    cin >> H >> W >> A >> B; 
    // Check if the grid can accommodate the required number of 0s and 1s
    if (W < 2 * A || H < 2 * B) { 
        cout << -1 << endl; // Print -1 if the grid is too small
        return 0; // Exit the program
    } 
    // Generate the grid based on the parameters A and B
    for (int i = 0; i < H; i++) { // Iterate over each row
        for (int j = 0; j < W; j++) { // Iterate over each column
            // Determine whether to write '0' or '1' based on the current position
            if ((i < B && j < A) || (i >= B && j >= A)) 
                grid[i][j] = '0'; // Write '0' for specified conditions
            else 
                grid[i][j] = '1'; // Write '1' otherwise
        } 
    } 
    // Print the grid
    for (int i = 0; i < H; i++) { // Iterate over each row
        for (int j = 0; j < W; j++) { // Iterate over each column
            cout << grid[i][j]; // Print the current character
        } 
        cout << endl; // Move to the next line after finishing a row
    } 
    return 0; 
} 
\end{lstlisting}
\vspace{-10pt}
\begin{lstlisting}[caption=Failed Test Case, label={lst:ex3-3}]
-- Input --
1000 1000 0 0
-- Expected Output -- A matrix of 1000 columns and 1000 rows with each value 0/1
000000000000000000.....
000000000000000000.....
.......................
-- Actual Output -- Error due to inappropriate grid variable size
\end{lstlisting}
The use of the word \textit{grid} in the comments in lines 11, 12, 13, 16, 18, and 21 of Listing \ref{lst:ex3-4} tricked the translating LLM into believing that \textit{grid} is a declared variable in the program. The translating LLM now becomes biased and declares a variable \textit{grid} (see Line 8, Listing \ref{lst:ex3-5}) with some randomly fixed dimensions (100*100). It continues to use this \textit{grid} variable to store the output matrix generated through the for loop in lines 17-25. It then prints the \textit{grid} using the for loop in lines 27-32. As long as the output matrix dimensions are less than (100*100), the program works fine. As soon as it exceeds this limit, there is a memory fault, and the corresponding test case fails. An example is given in Listing \ref{lst:ex3-3}. This example precisely demonstrates how translating LLMs are negatively impacted by the choice of words in code comments.

\subsection{RQ2 : Impact of comment intent on translation performance}
\label{subsec:rq2}
\vspace{-20pt}
\begin{table}
  \centering
  \includegraphics[scale=0.5]{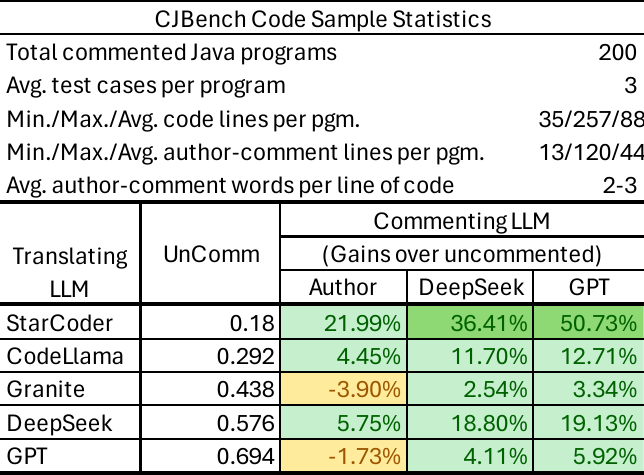}
  \vspace{5pt}
  \caption{CJBench Statistics and J2P Translation Results
  }
  \label{tab:cjbenchresults}
\end{table}
\vspace{-30pt}
Code comments can be model-generated, usually short and descriptive, or author-written, which can exhibit several intents, such as explaining the algorithm, the code, its complexity, usage, etc.
To study the impact of comment intents on the translation performance of LLMs, we created a dataset : \emph{CJBench} - \textbf{C}ommented \textbf{J}ava Programs \textbf{Bench}mark (Table \ref{tab:cjbenchresults}) - a manually curated collection of 200 author-commented Java programs, sourced from (a) Java-Projects-Collection, a collection of various Java-Projects\footnote{\url{https://github.com/kishanrajput23/Java-Projects-Collections}} and (b) \textit{The Algorithms}, a large open-source algorithm library\footnote{\url{https://github.com/TheAlgorithms/Java/}}.

\begin{figure}[t]
\centering
  \includegraphics[scale=0.4]{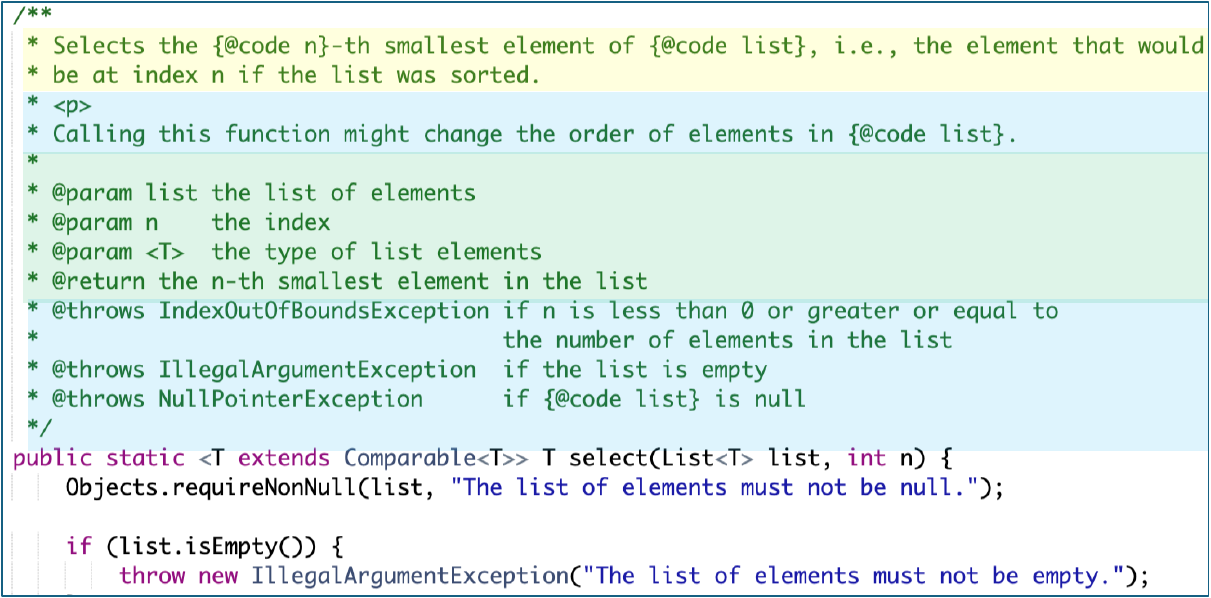}
  \caption{Example of one big multi-line (12 lines) comment with several intents}
 \label{fig:comment}
 \includegraphics[width=\linewidth]
  {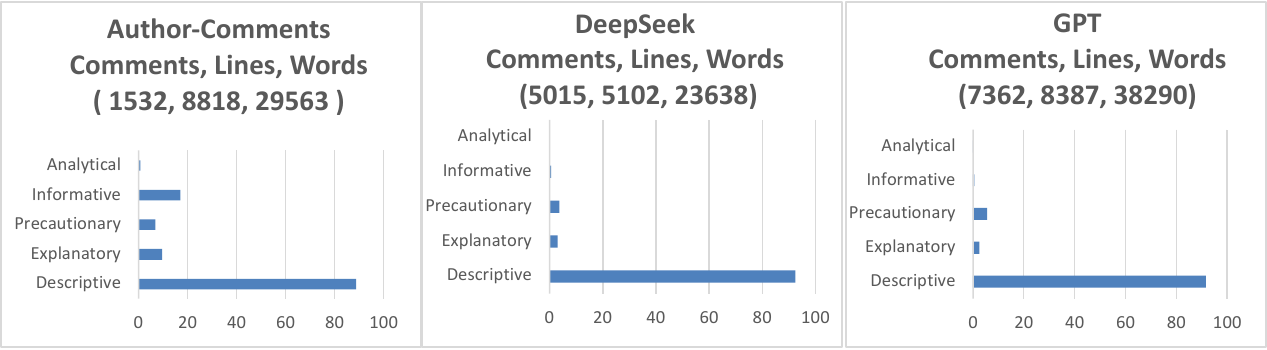}
  \caption{Comparison of Intent Classes in Comments}
  \label{fig:intent}
\end{figure}

\noindent\textbf{Are multi-line, multi-intent comments useful in code translation?} Figure \ref{fig:comment} presents an example of an elaborate multi-line, multi-intent comment extracted from a code sample in the CJBench dataset. It \textit{\textbf{describes}} (highlighted in yellow) the objective of the associated method, \textit{\textbf{cautions}} (highlighted in blue) that the method can change the order of elements in the list, and also the exceptions it can throw, and \textit{\textbf{informs}}
(highlighted in green) the list of I/O parameters. To understand how these multi-intent comments  affect code translations, we performed Java to Python (J2P) translations of code samples in the CJBench dataset. For comparison 
with model-generated comments, which are comparatively small and descriptive,  we first removed the original author-comments, and then DeepSeek and GPT models were used to inject comments into the code samples. Table \ref{tab:cjbenchresults} presents the comparative translation performance on these three datasets. As expected and clearly seen from the low/negative gain numbers (in yellow), author-commented samples performed poorer than DeepSeek-commented or GPT-commented samples. To further establish the reason behind this comparatively poorer performance, we conducted an extensive intent-classification experiment.

\noindent\textbf{Intent classification.} All author and model generated comments in the CJBench code samples were extracted, and GPT was instructed to perform intent classification on the 50 largest comments. The resultant intent categories were: 
\begin{itemize}[noitemsep,topsep=2pt]
    \item \textbf{Descriptive} - describe what the code does.
    \item \textbf{Explanatory} - explain the overall approach used.
    \item \textbf{Informative} - specify details about the I/O parameters.
    \item \textbf{Analytical} - mention the performance details.
    \item \textbf{Precautionary} - warn about  potential risks/exceptions.
\end{itemize}
 Subsequently, GPT was used to classify the remaining comments into one or more of these categories (see the 
 Appendix
 for examples).\\
\noindent\textbf{Which comment intent is most useful in code translation?} From Figure \ref{fig:intent}, it can be calculated that the average number of words in an author-comment is 29563/1532 = 19, while this value is around 5 for model-comments. In other words, each author-comment typically contains more words and lines than model-comments, as it touches on several aspects of the corresponding code (see Figure \ref{fig:comment}).
Figure \ref{fig:intent} proves this claim. While author-comments do have a Descriptive intent; however, unlike model-generated comments, there is also a prominent presence of Explanatory, Precautionary, and Informative intents in them. Looking at the gain numbers in Table \ref{tab:cjbenchresults}, we see that the author-comments result in nearly half the gain compared to the model-comments. The only reason behind this could be the presence of multiple other intents alongside the only necessary - Descriptive type -  intent. Perhaps the close presence of other intents is creating noise that the translating LLM is unable to cancel out, resulting in comparatively poorer performance. 

Key Takeaway : Short comments that \textit{describe} what the code does are more useful to LLMs during code translation. Other comment types may add to noise. 
\begin{figure}[t]
\centering
\includegraphics[width=\linewidth]{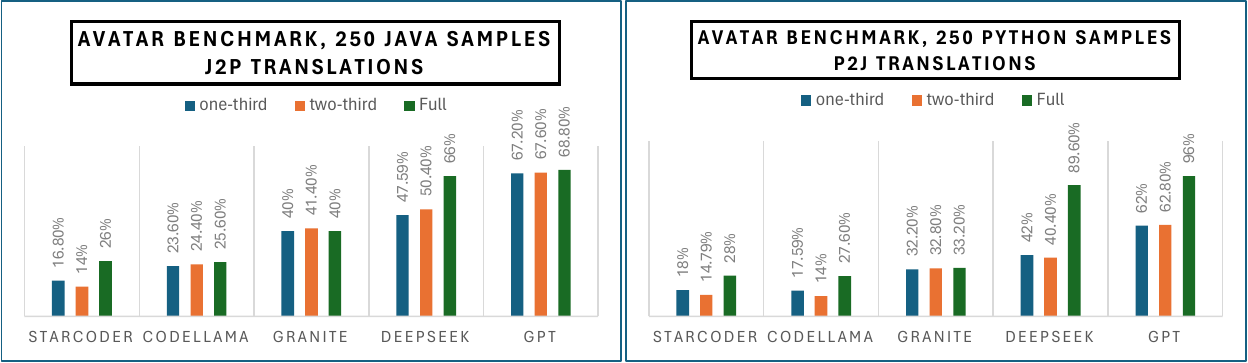}
  \caption{Impact of Comment Density on Translation}
  \label{fig:densityimg}
\end{figure}
\begin{figure}[t]
\centering
\includegraphics[width=\linewidth]{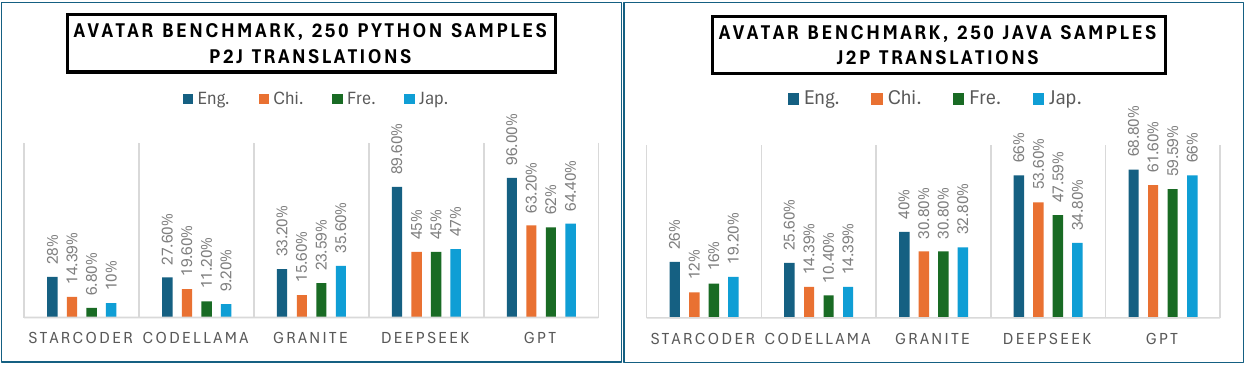}
  \caption{Impact of Comment Language on Translation}
  \label{tab:lang}
\end{figure}
\vspace{-22pt}
\subsection{RQ3 : Density and Language of Comments}
\label{subsec:rq3}
\vspace{-5pt}
\noindent{\textbf{Impact of Comment Density.}} We noted in the previous two studies that, although comments are helpful in code translation, they tend to add noise to the code being translated. In an attempt to shed light on how much code should be commented on without adding too much noise while still being useful for code translation, we conducted the following experiment.
In 250 Java and 250 Python code samples from the AVATAR dataset, we instructed the commenting LLM - GPT, to restrict the inclusion of code comments to only the top one-third and two-thirds of the complex lines of code instead of the entire code. The commenting LLM was instructed to decide for itself which lines of code to comment. Usual J2P and P2J translations were then attempted on these differently commented samples using various translating LLMs, and the percentage of successful translations was recorded. The results are summarized in Figure \ref{fig:densityimg}. 
It is evident that limiting the density of comments did not prove beneficial. One could argue that this may be due to uncertainty about which sections of the code would be commented on when such restrictions are imposed. While this is valid, identifying which code blocks require comments remains a challenge until an error-to-code block mapping is obtained for the failed translation samples. 

Key Takeaway : Arbitrary restrictions on comment density did not prove beneficial. More experiments may be needed to understand the role of selective commenting based on heuristics or error-localization–driven selection.\\
\noindent{\textbf{Is English as a commenting language better than other languages? }}
To study whether the choice of natural language (English, Japanese, French, Chinese, etc.) for code comments impacts the code translation performance of LLMs, we conducted the following experiment. In 250 Java and 250 Python code samples from the AVATAR dataset, the English language comments generated by the GPT model were translated into 3 other natural languages: Japanese, French, and Chinese, using the GPT model itself. While generating comments natively in each of these 3 other natural languages using specialized multilingual code models was another option, for this experiment, we chose to translate English comments into these languages to avoid any potential bias in results comparison due to the introduction of new commenting models. J2P and P2J translations of these different language commented sets were then attempted using various translating LLMs. The results are presented in Figure \ref{tab:lang}. We observe that in almost all cases, English comments performed the best for both J2P and P2J translation tasks, the only exception being P2J translations on the Granite model, which show marginally better performance with Japanese comments. 
\begin{figure}[t]
\centering
\includegraphics[width=\linewidth]{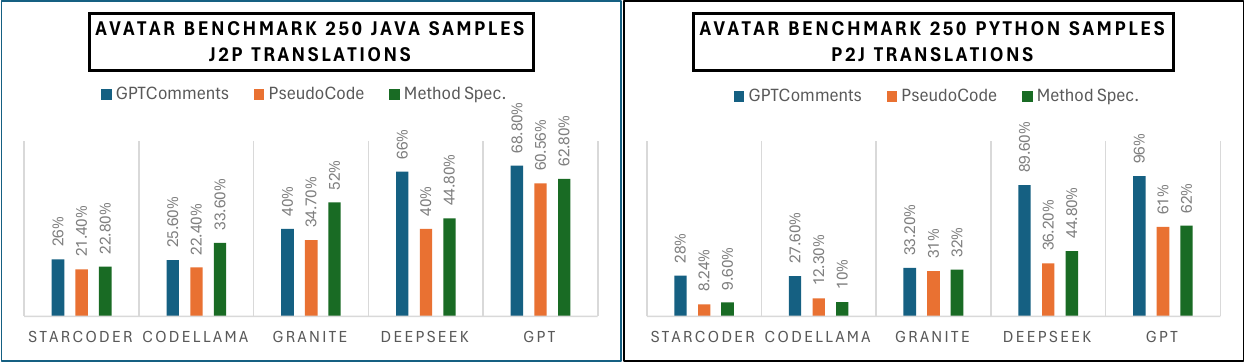}
  \caption{ Impact of Location on Translation : Code Comments vs. Method Specifications \& Pseudo-code}    
  \label{tab:placement}
\end{figure}
\vspace{-8pt}
\subsection{RQ4 : Impact of Code Comments vs. Method Specifications \& Pseudo-code}
\label{subsec:rq4}
\vspace{-5pt}
We now attempt to answer another important question about code comments: How does the placement of code specifications affect translation performance? We explore the impact of placing NL code specifications in 3 different forms at 3 different locations: (1) in the form of \textit{pseudocode} , which will be appended to the source code in the translation prompt, (2) in the form of \textit{method specifications}, placed at the beginning of each method within the source code, and (3) in the form of code comments, closely placed alongside lines-of-code. We used our commenting LLM, GPT, to generate pseudocode and method specifications (one for each method in the code sample) for a total of 500 (250 Java and 250 Python) AVATAR code samples. Pseudocodes were appended to the source code in the translation prompt, while method specifications were placed at the beginning of each method within the source code. J2P and P2J translation results of these code samples using various translating LLMs are presented in
Figure \ref{tab:placement}. We observe that NL-specifications in the form of pseudocode, placed in the translation prompt, were definitely not as promising as method specifications or code comments, both of which were placed within the code itself. Apart from J2P translations on CodeLlama and Granite, for all other cases, the code comments approach yielded higher translation performance compared to the method specifications approach.

Key Takeaway : Code comments are more effective than method specifications or pseudocode in enhancing the translation performance of LLMs.

\begin{table}[t]
\centering
  \includegraphics[width=\textwidth]{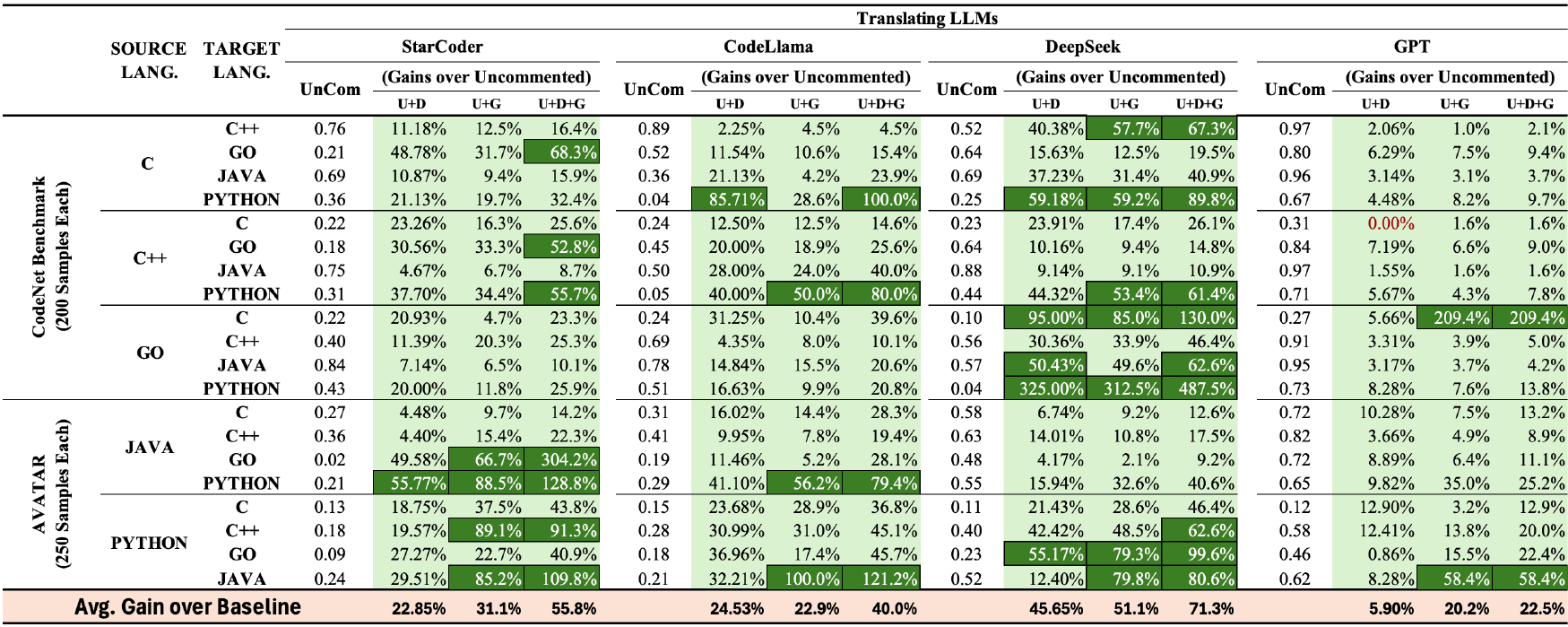}
  \caption{COMMENTRA - Gains in translation performance of various LLMs on different input samples. Dark green cells show the best performing configuration. Results are shown for a subset of translating LLMs. Results for all translating LLMs are in 
  Appendix.
  }
  \vspace{-10pt}
  \label{commentra-res}
\end{table}
\vspace{-6pt}
\section{The COMMENTRA Approach}
\begin{figure}[t]
\centering
  \includegraphics[scale=0.3]{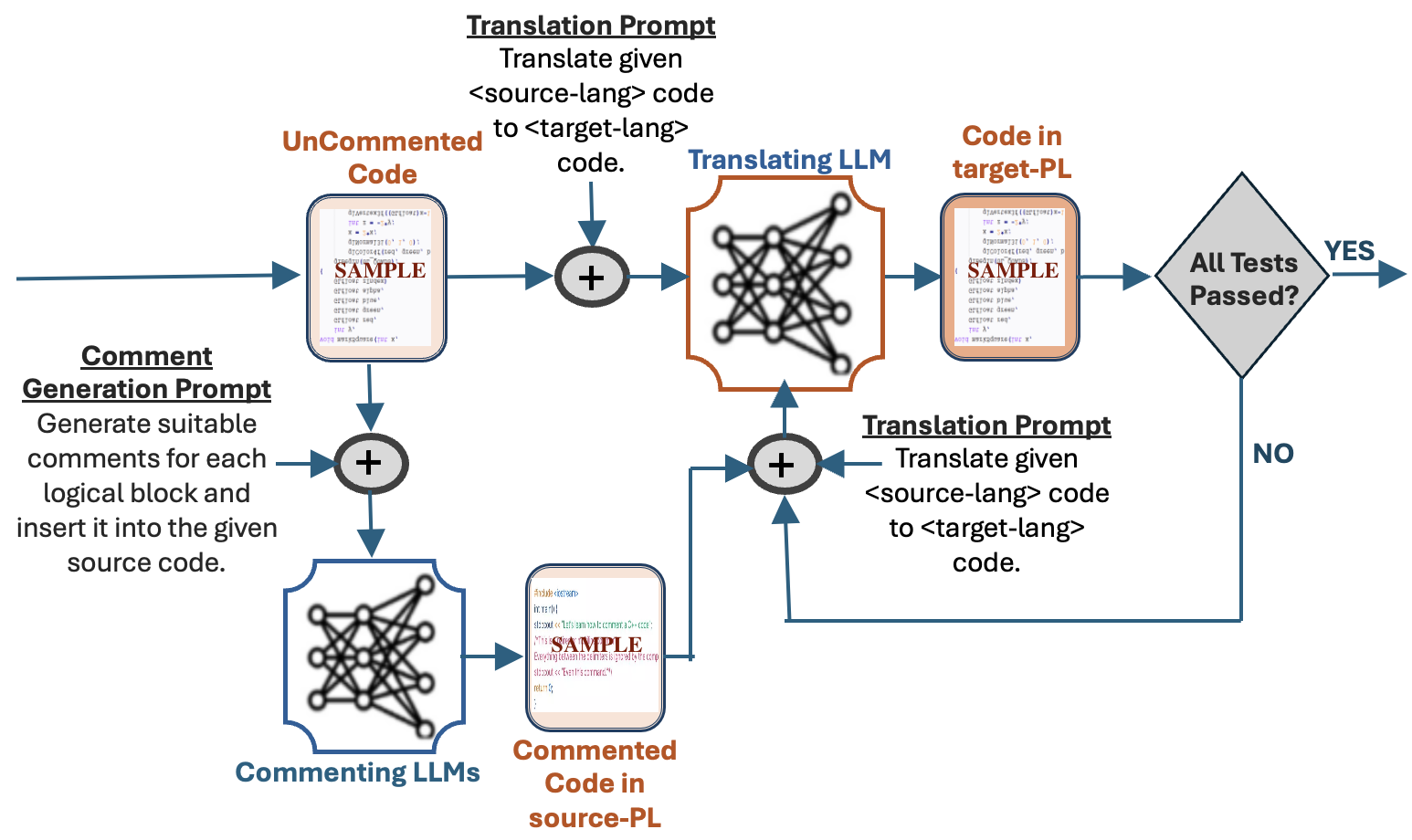}
  \caption{COMMENTRA: Comment-Based Translation}
  \label{commentra-set}
\end{figure}

\vspace{-6pt}
Based on the results and key takeaways from our study, we now propose a comment-based iterative translation approach - \textit{COMMENTRA}. 
The central insight underpinning COMMENTRA is that, although comments can significantly improve the code translation capabilities of LLMs, their indiscriminate use may be counterproductive. Consequently, comments are incorporated only when initial translations without comments prove unsuccessful. This targeted approach delivers both time and cost efficiencies. By selectively introducing comments and reattempting failed translations, COMMENTRA achieves substantial improvements, resolving a wide range of test case failures as well as runtime and compilation errors. Figure \ref{commentra-set} presents the setup of the proposed approach. The translation process works in iterations.  In the initial iteration, translations of all uncommented code samples in source-PL are attempted. The output code samples in the target-PL are then tested for errors. All input samples for which the translation is erroneous move to the second iteration, wherein comments are first injected into the corresponding original source code files using a suitable commenting model, and then the translation is reattempted. Multiple iterations with different commenting models can be performed in a similar way, depending on the desired accuracy and the budget available for model invocations. As long as we follow the iterative approach suggested by COMMENTRA, performance consistently improves. 

Table \ref{commentra-res} shows the gains in translation performance achieved by the COMMENTRA approach with up to three iterations. For each translating LLM, the first iteration is the translation of uncommented code samples. The second iteration uses DeepSeek (labeled as \textbf{U+D} in the results table) or GPT (labeled as \textbf{U+G}) as the commenting model. 
 The column labeled \textbf{U+D+G} shows the gains in translation performance after three iterations: iteration 1 of uncommented code samples, iteration 2 of DeepSeek-commented code samples\footnote{DeepSeek is used to inject comments only in those uncommented code samples that failed to translate correctly in the first iteration.}, and iteration 3 of GPT-commented code samples\footnote{GPT is used to inject comments only in those uncommented code samples that failed to translate correctly even after the first and second iterations.}. As GPT is a paid model, a series of three iterations \textbf{U+D+G} may be performed in a cascade to reduce costs, wherein the maximum benefit is first derived from open source models such as DeepSeek, and then GPT is used only for the failed translation samples after iteration 2. The bold entries in the table correspond to the cases where the gain in translation performance is more than 50\% after running through the two or three iterations of COMMENTRA. In our experiments, for all pairs of languages and LLMs that translated / commented, the average gain in translation performance achieved by one iteration of the commented code ranged between 6\% and 51\%. The same after two commented code iterations ranged from 23\% to 71\%.

To further validate COMMENTRA, we conducted additional experiments on the CodeTransOcean benchmark; detailed results are in the 
Appendix.

\vspace{-12pt}
\section{Threats to Validity}
\vspace{-8pt}
We conducted our study on standalone programs in 5 PLs. 
However, more research is needed to assess the impact of comments in the translation of other language pairs, as well as the impact of code comments on large-scale, real-world code repositories or complex codes. In the case of complex codes, measuring and comparing the quality of comments generated by LLMs will be important.

Our observations reveal that  Mistral-generated comments did not result in a significant improvement in translation performance. Similarly,  StarCoder and CodeLlama, as translation models, did not benefit much from the inclusion of code comments. However, it is important to interpret these numerical results with some reservations, as most of the models we examined are still evolving and improving in capability. Additionally, since the inclusion of comments contributes to a significant increase in contextual information, translating very large programs with comments can pose challenges due to LLM's token length constraints. One potential approach to mitigate this issue is to selectively add comments based on code complexity or error-localization analysis.

\vspace{-10pt}
\section{Conclusions}
\vspace{-8pt}
We explore the use of code comments as contextual information in the source code. Our experiments span multiple models and PLs, 
and we present a comprehensive study on the impact of code comments on code translation, considering multiple perspectives such as density, language, intent, and placement of comments. Our results demonstrate that while comments can enhance translation performance, they can also introduce noise and degrade translations. Building on all our insights, we propose COMMENTRA, an iterative translation approach to translating code with comments when the raw translation fails. 
Experimental results show that this simple-to-implement yet highly effective approach can substantially improve LLMs translation performance.

\section{Acknowledgment}
All authors acknowledge the support of the IBM AI Horizons Network at IIT Delhi for this work. Srikanta Bedathur acknowledges his DS Chair in AI professorship at IIT Delhi. 


\clearpage
\appendix
\section{Related Work Comparison}
\begin{table}[hbt!]
  \centering 
  \resizebox{\textwidth}{!}{
  \begin{tabular}{lp{15cm}C{4cm}p{6cm}}
    \hline
    \rowcolor{lightgray!25}
    \textbf{Related Work} & \textbf{Idea} & \textbf{Benchmarks Used} & \textbf{Comparison with COMMENTRA} \\
    \hline
    
    \\
    \textbf{\emph{UniTranslator}} \cite{unitranslator} &
    {\raggedright Proposes a dynamic, multiple LLM framework for code translation, wherein specialized LLMs, each with unique expertise, are orchestrated by a Director LLM to achieve high-fidelity translations and proactively address potential bugs.} &
    { CodeNet, AVATAR} &
    {\raggedright Our approach also suggests using multiple LLMs - specialized Commenting LLMs and Translating LLMs.} \\
    \\
    
    \hline
    \rowcolor{lightgray!25}
    \multicolumn{4}{c}{\textbf{\textit{Approaches suggesting augmented source-code and/or code-translation-prompt}}} \\
    \hline

    \\
    \textbf{\emph{Saha et al.}} \cite{saha} &
    {\raggedright Investigates utility of NL-specifications as an intermediate representation for code translation. Their results show that using NL-specifications combined with source code, provides a slight improvement over the baseline in certain language pairs.} & 
    { CodeNet, AVATAR, Evalplus} & 
    \multirow{4}{=}{{\raggedright Our approach suggests using code-comments to augment source-code in cases where the original uncommented source-code translation fails. Code comments differ from NL-specifications in terms of their close, precise association and presence with the corresponding line(s)-of-code. While NL-specifications explain the overall code logic, code comments, describe the associated line(s)-of-code. Also code comments can be human-written or model-generated and can exhibit several intents. None of these work do a systematic investigation of how comments (positively and negatively) impact code translation.}} \\

    \\
    \textbf{\emph{Tang et al.}} \cite{tang} &
    {\raggedright Explored the use of self-generated
natural language explanations as an intermediate step for code-to-code translation. Their results show that some improvements with natural language explanations which are particularly pronounced on difficult programs.} & 
    { MultiPL-C2C} & 
    \\

     \\
    \textbf{\emph{SpecTra}} \cite{spectra} & 
    {\raggedright SpecTra first generates high-quality static specifications, test cases, and natural language descriptions from a given source code, and then uses them to augment original source code, with the aim to improve the overall translation performance of LLMs.} &
    { CodeNet }& \\
     
    \\
    \textbf{\emph{UniTrans}} \cite{unitrans} & 
    {\raggedright UniTrans first leverages LLMs to auto-generate a series of test cases for target programs with the assistance of source programs. Next, it harnesses the above test cases to augment the code translation prompt. The translated programs are double-checked for correctness via test-cases execution. The incorrectly translated programs are iteratively repaired in multiple rounds based on feedback from test cases.} & 
    { Standalone code samples from \emph{GeeksforGeeks} online platform} & \\
     
\\
    \hline
    \rowcolor{lightgray!25}
    \multicolumn{4}{c}{\textbf{\textit{Translation repair approaches}}} \\
    \hline

    \\
    \textbf{\emph{LANTERN}} \cite{lantern} &
    {\raggedright Suggests a program repair approach, wherein, when an LLM fails to repair buggy code in a given PL, the bug is strategically translated into another alternate PL, where the model
    demonstrates stronger repair capabilities. The alternate PL is selected based on bug characteristics and historical feedback. Once the bug is translated, repair is attempted in 
    this alternative language, and the successfully repaired code is subsequently translated back to the original language.}  & 
    { xCodeEval} & 
    \multirow{4}{=}{{\raggedright Our bug-repair approach is comparatively low-cost and can be easily integrated with all these approaches to quick-fix a considerable number of translation bugs. In future, we can train Commenting LLMs to generate code-comments that specifically help Translating LLMs in preventing common code translation bugs.}} \\
    
    \\
    \textbf{\emph{FLOURINE}} \cite{flourine} &
    {\raggedright The proposed FLOURINE framework first prompts the candidate LLM to generate the RUST language translation of the input source program, and if the translation does not compile, it is prompted to repair compilation errors. Once the translation compiles, a cross-language differential fuzzer is invoked, which tests the translation for equivalence to the original, and returns counterexamples if the translation is not equivalent. The counterexample is used as input to repair the buggy translation. Once the fuzzer fails to find a counterexample after a configured timeout, the translation is considered validated.} & 
    { Standalone code samples from real-world projects that can fit in the context  window of LLMs and that use only standard libraries.} & \\
    
    \\
    \textbf{\emph{Rectifier}} \cite{rectifier} &
     {\raggedright Proposes a general corrector, namely Rectifier, which is a micro and universal model for repairing translation errors. It learns from errors generated by existing LLMs and can be widely applied to correct errors generated by any LLM. } &
     { CodeNet, AVATAR} & \\

    \\
    \textbf{\emph{TRANSAGENT}} \cite{transagent} &
     {\raggedright The unique feature of TRANSAGENT approach is to divide the source program into blocks based on the control flow graph, and then leverage LLMs to map each block of the source program to that of the target program. Subsequently based on mapped blocks and execution
     alignment, the error code block is identified in the target program. LLMs are then used 
     to specifically fix the identified error block with the observed runtime difference.} &
     { Standalone code samples from \emph{GeeksforGeeks} and \emph{LeetCode} online platform} & \\
    \\
     
    \hline
    \rowcolor{lightgray!25}
    \multicolumn{4}{c}{\textbf{\textit{Intermediate language translations}}} \\
    \hline
    \\
    \textbf{\emph{INTERTRANS}} \cite{intertrans} &
     {\raggedright Suggests leveraging intermediate translations to bridge the syntactic and semantic gaps between source and target PLs. A translation tree containing all potential translation paths for the  specific source-target PL pair is first generated. Translation paths are then turned into LLM prompts that are executed in a breadth-first order. Available test suites are used to validate whether the generated target translation is correct, enabling early termination of translation path exploration if a successful path is found before completely exploring the translation tree.} &
    { CodeNet, HumanEval-X, TransCoder} &
    \multirow{3}{=}{{\raggedright All intermediate-level code-translations to another PL, can potentially deviate from the original source code and introduce new translation bugs. In contrast, in our approach, source-code augmented with code-comments is still closer to the original source-code and therefore may not potentially introduce new translation-bugs.}} \\

    \\
    \textbf{\emph{Szafraniec et al.}}\cite{szaf} &
    {\raggedright Proposed to augment code translation with low-level compiler intermediate representations (IRs).}& 
    { Projects extracted with Google BigQuery} & \\
    \\

    \\
    \textbf{\emph{Tao et al.}}\cite{llmpotential} &
    {\raggedright The work reports LLMs' suboptimal code translation performance on Python to other PLs and the negligible impact of widely adopted LLM optimization techniques such as conventional pre-training and instruction tuning on code translation. Further, to enhance code translation performance, this work proposes selecting an intermediary translating language between the source and target PL and fine-tuning LLMs on self-generated parallel data.}& 
    { PolyHumanEval} & \\
    \\
    
    \hline
    \rowcolor{lightgray!25}
    \multicolumn{4}{c}{\textbf{\textit{LLM fine-tuning approaches}}} \\
     \hline
     \\
     \textbf{\emph{CoTran}} \cite{jana} &
     {\raggedright CoTran suggests fine-tuning code translation LLMs using reinforcement learning, feedback from compiler and symbolic-execution-based testing feedback to assess functional equivalence between the input and output programs.} & 
    { AVATAR-TC} & \\
    \\
    
    \hline
    \rowcolor{lightgray!25}
    \multicolumn{4}{c}{\textbf{\textit{Repository-level and Class-level code translation}}} \\
    \hline
    
    \\
    \textbf{\emph{AlphaTrans}} \cite{alphatrans} &
     {\raggedright AlphaTrans leverages program analysis to decompose repository-level programs into fragments and attempts translating them in the reverse call order.} &
     { Real-world open-source projects} &
     \multirow{3}{=}{{\raggedright Our work currently focuses on understanding the utility of code comments in the general code-translation process}}\\
     
    \\
   \textbf{\emph{RepoTransBench}} \cite{repotransbench} &
    {\raggedright Proposes a new benchmark, RepoTransBench, which is a real-world repository-level code translation benchmark with an automatically executable test suite.} & 
    { RepoTransBench} & \\
    \\

    \textbf{\emph{ClassEval-T}} \cite{classevalt} &
    {\raggedright Proposes a new class-level benchmark ClassEval-T, and demonstrated that LLMs perform much worse on class-level code translations.} & 
    { ClassEval-T} & \\
    \\
    
    \hline
    \rowcolor{lightgray!25}
    \multicolumn{4}{c}{\textbf{\textit{Code comment generation}}} \\
     \hline

     \textbf{\emph{DOME}} \cite{mu2023developerintentdrivencodecomment, mu} &
      {\raggedright A piece of code can be commented with several intents such as its functionality, its usage or its space-time complexity etc.. These works propose an intent-driven code comment generation approach DOME, to produce a comment coherent with the given intent.} & &
      {\raggedright Currently, our Commenting LLMs mostly generate comments with \textit{Descriptive} intent. In future, we can explore generation of code-comments with \textit{ CodeTranslation} intent.} \\
      
    \hline
  \end{tabular}
 }
  \label{tab:related-work-comparison}
\end{table}
\vspace{-20pt}
\begin{table}
  \centering
  \resizebox{\textwidth}{!}{%
  \begin{tabular}{l|c|c|c|c}
    \hline 
    \\
    \textbf{Related Work} & 
    {\raggedright{\textbf{Need multiple (specialized) LLMs}}} & 
    {\raggedright{\textbf{Use NL specifications to augment source code to be translated}}} & 
    {\raggedright{\textbf{Use NL specifications/comments to train models}}} & 
    {\raggedright{\textbf{Same/Super-set/Subset of Benchmarks Used For Experimentation}}} \\ \\
    \hline 
    \\ 
    \textbf{\emph{COMMENTRA}} &
    \ding{52} &
    {\ding{52} Suggests calibrated use of \textbf{Code Comments}} &
    \ding{56} &
    \ding{52} \\
    \\
    
    \textbf{\emph{UniTranslator}} \cite{unitranslator} &
    \ding{52} &
    \ding{56} &
    \ding{56} &
    \ding{52} \\
    \\
    
    \textbf{\emph{Saha et al.}} \cite{saha} &
    \ding{56} &
    {\ding{52} Investigates use of \textbf{NL-specifications}} &
    \ding{56} &
    \ding{52} \\
    \\

    \textbf{\emph{Tang et al.}} \cite{tang} &
    \ding{56} &
    {\ding{52}  Explore the use of \textbf{self-generated natural language explanations}} &
    \ding{56} &
    \ding{56} \\
    \\
    
    \textbf{\emph{SpecTra}} \cite{spectra} & 
    \ding{56} &
    {\ding{52} Explore the use of \textbf{static specifications, test cases, and natural language descriptions}} &
    \ding{56} &
    \ding{52} \\
    \\

    \textbf{\emph{Du et al.}} \cite{dua} & 
    \ding{56} &
    \ding{56} &
    {\ding{52} Suggest use of code comments to train models} &
    \ding{52} \\
    \\

    \textbf{\emph{TransCoder}} \cite{roziere2020unsupervised}  & 
    \ding{56} &
    \ding{56} &
    {\ding{52} Suggest use of code comments to train models} &
    \ding{52} \\
    \\

    \textbf{\emph{SpecTra}} \cite{spectra} & 
    \ding{56} &
    {\ding{52} Explore the use of \textbf{static specifications, test cases, and natural language descriptions}} &
    \ding{56} &
    \ding{52} \\
    \\
     
    \textbf{\emph{UniTrans}} \cite{unitrans} & 
    \ding{52} &
    \ding{56} &
    \ding{56} &
    \ding{56} \\
    \\
    
    \textbf{\emph{LANTERN}} \cite{lantern} &
    \ding{56} &
    \ding{56} &
    \ding{56} &
    \ding{56} \\
    \\
    
    \textbf{\emph{FLOURINE}} \cite{flourine} &
    \ding{56} &
    \ding{56} &
    \ding{56} &
    \ding{56} \\
    \\
    
    \textbf{\emph{Rectifier}} \cite{rectifier} &
    \ding{52} &
    \ding{56} &
    \ding{56} &
    \ding{52} \\
    \\
    
    \textbf{\emph{TRANSAGENT}} \cite{transagent} &
    \ding{56} &
    \ding{56} &
    \ding{56} &
    \ding{56} \\
    \\
    
    \textbf{\emph{INTERTRANS}} \cite{intertrans} &
     \ding{56} &
    \ding{56} &
    \ding{56} &
    \ding{52} \\
    \\
    
    \textbf{\emph{Tao et al.}}\cite{llmpotential} &
    \ding{56} &
    \ding{56} &
    \ding{56} &
    \ding{56} \\
    \\
    
     \textbf{\emph{CoTran}} \cite{jana} &
    \ding{56} &
    \ding{56} &
    \ding{56} &
    \ding{56} \\
    \\
    \hline
  \end{tabular}%
  }
  \label{tab:related-work-check-comparison}
  \caption{An overview comparison of our work and the various features of our approach with existing research works}
\end{table}
\clearpage
\section{RQ1 : Full Results of Table 2}
\begin{figure}
\centering
\includegraphics[width=\columnwidth]{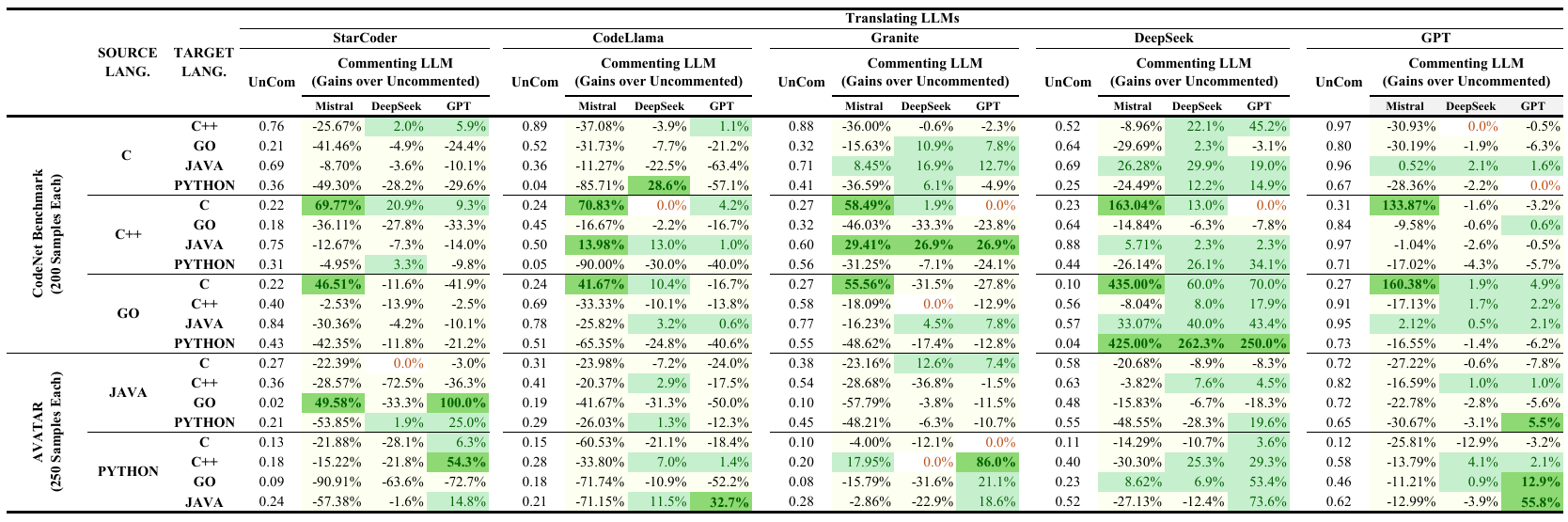}
  \caption{Full results : Impact of comments generated by various commenting LLMs over
baseline translation performance (accuracy) of different translating LLMs across
different pairwise language translations in benchmark datasets.}
  \label{fig:rq1-table-full}
\end{figure}
\clearpage
\section{Author/Model Comments Classification Examples}
Presented below are some examples from \textit{CJBench} of Author-written comments and Model-generated comments along with their intent classification results.
\vspace{5pt}

\begin{figure*}[h!]
\begin{subfigure}{.5\textwidth}
  \centering
    \fbox{\includegraphics[width=0.94\linewidth]{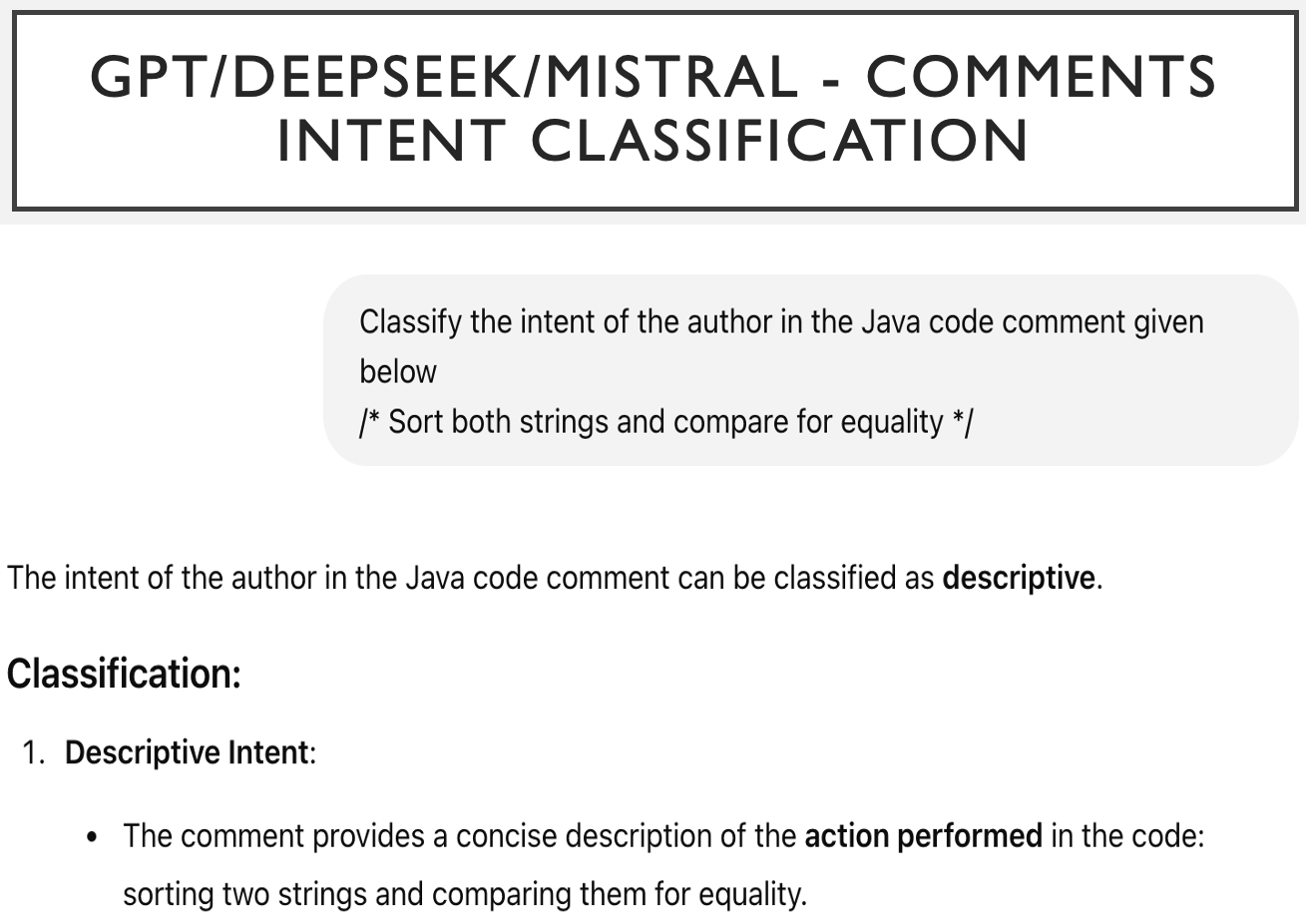}}
  \end{subfigure}
\begin{subfigure}{.5\textwidth}
  \centering
    \fbox{\includegraphics[width=0.95\linewidth]{ 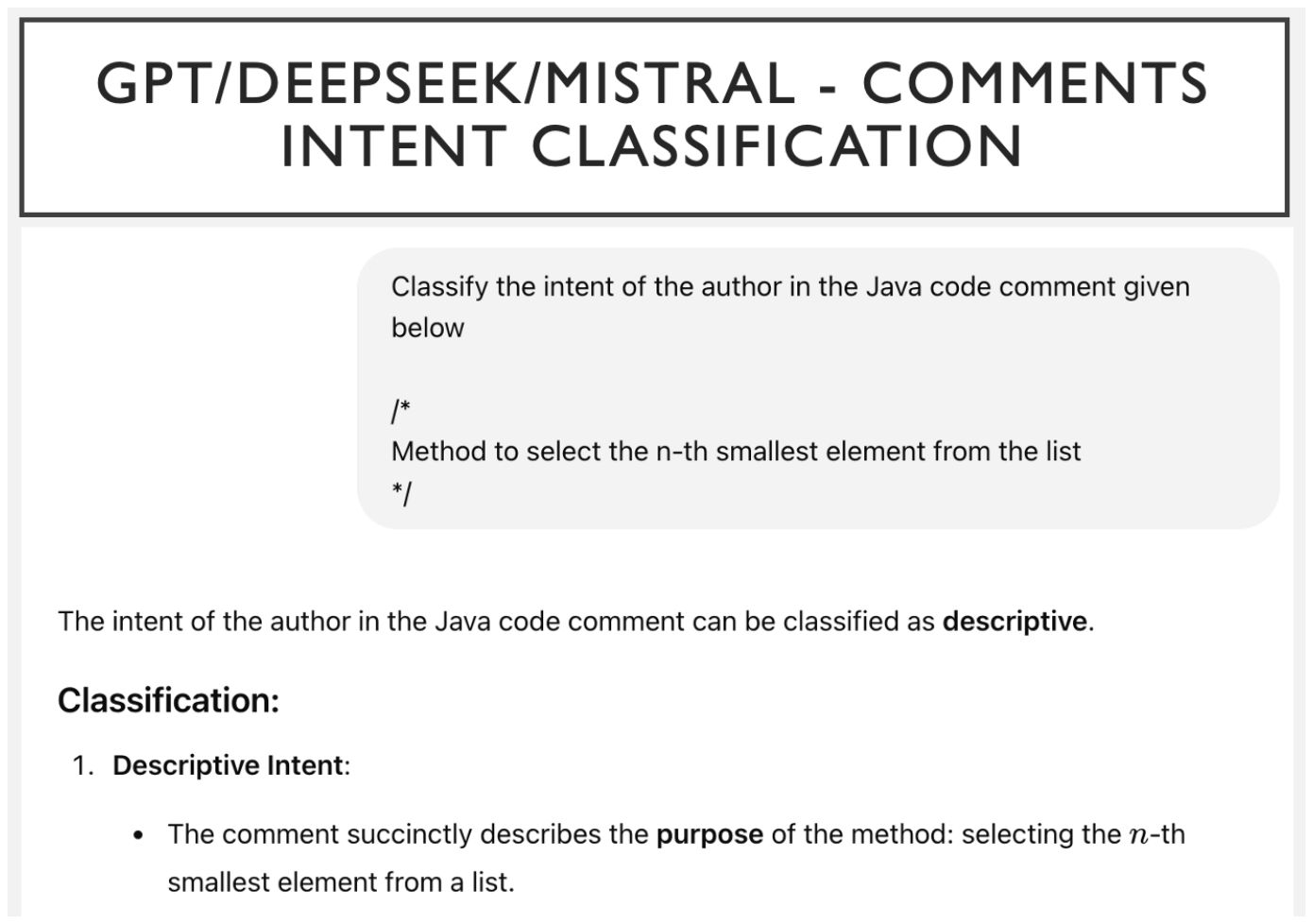}}
  \end{subfigure}
\end{figure*}

\begin{figure*}[h!]
\centering
\fbox{\includegraphics[width=0.8\linewidth]{ 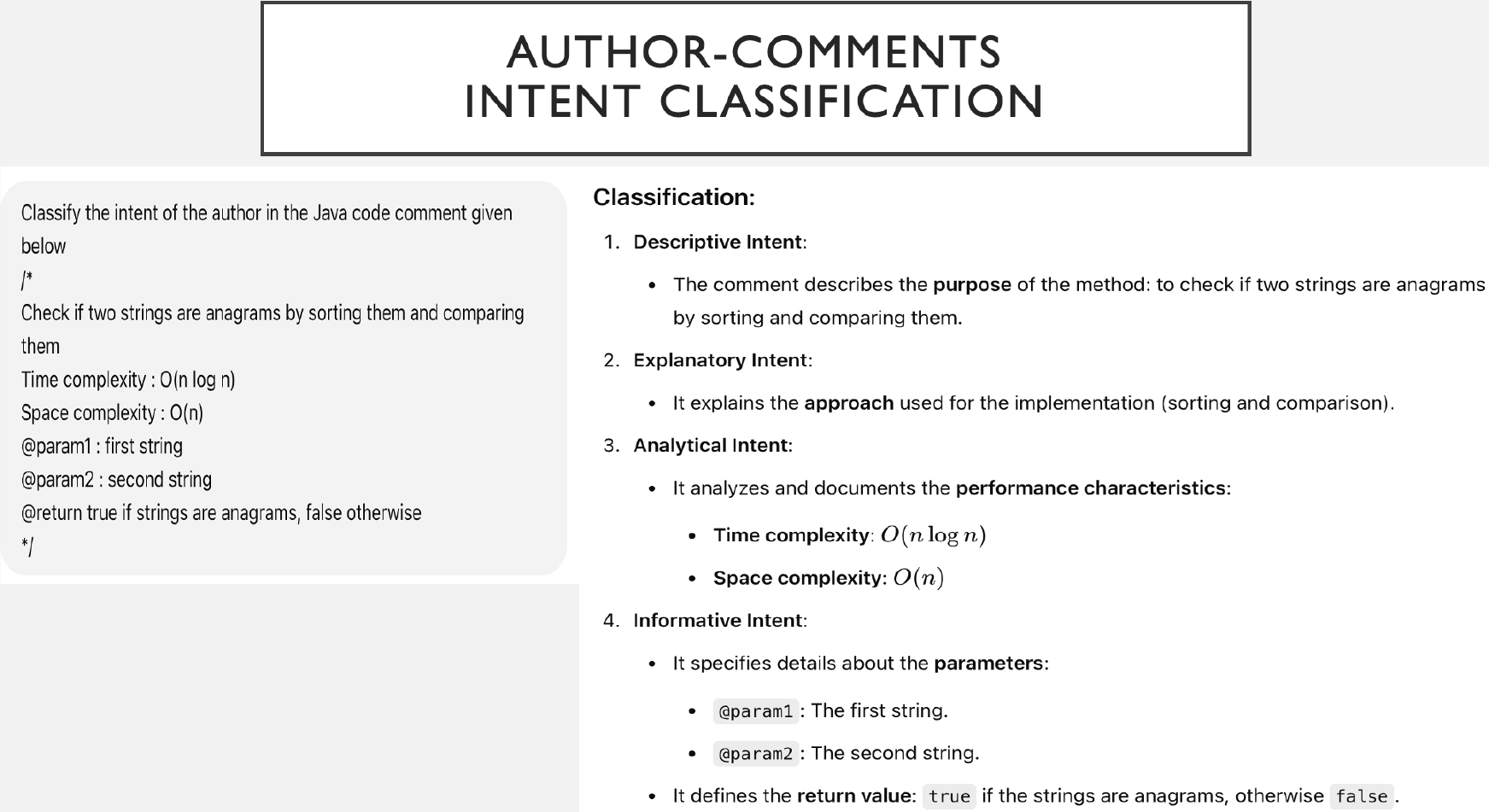}}
\end{figure*}
\clearpage
\section{COMMENTRA - Additional Results}
\subsection{COMMENTRA : Full Results of Table 4}
\begin{figure}
\centering
\includegraphics[width=\columnwidth]{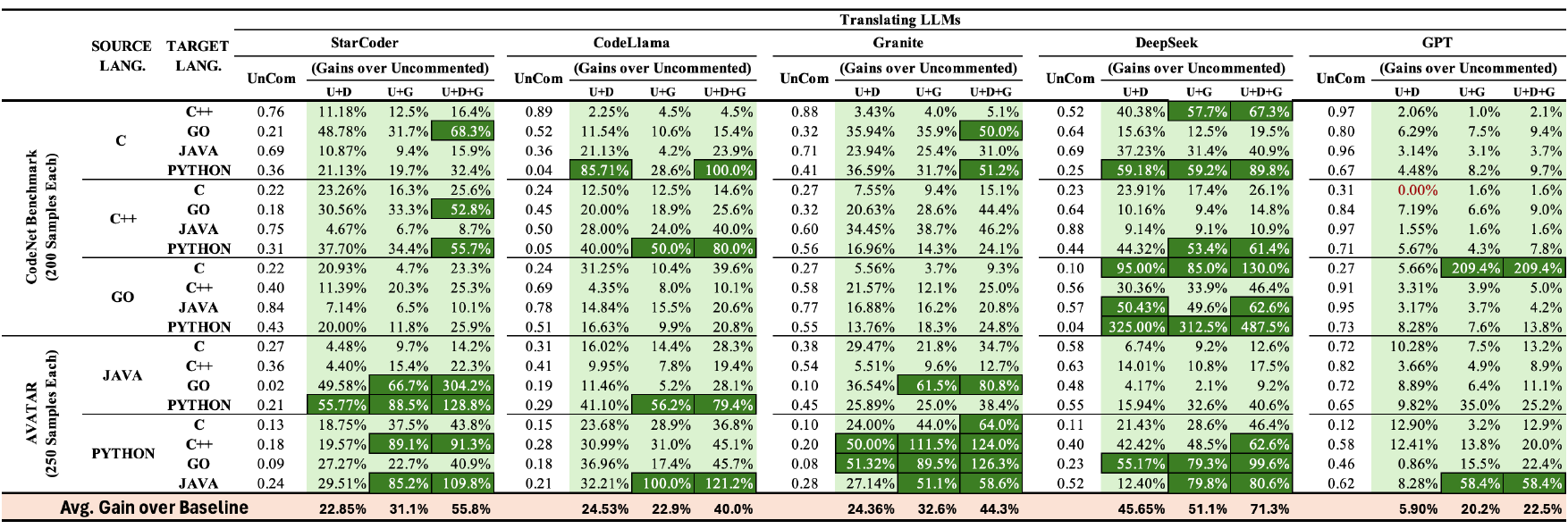}
  \caption{Full results : COMMENTRA - Gains in translation performance of various LLMs on
different input samples. Dark green cells show the best performing configuration. }
  \label{fig:rq1-table-full}
\end{figure}
\subsection{COMMENTRA : Results of CodeTransOcean Benchmark}
\begin{figure}
\centering
\includegraphics[width=\columnwidth]{
 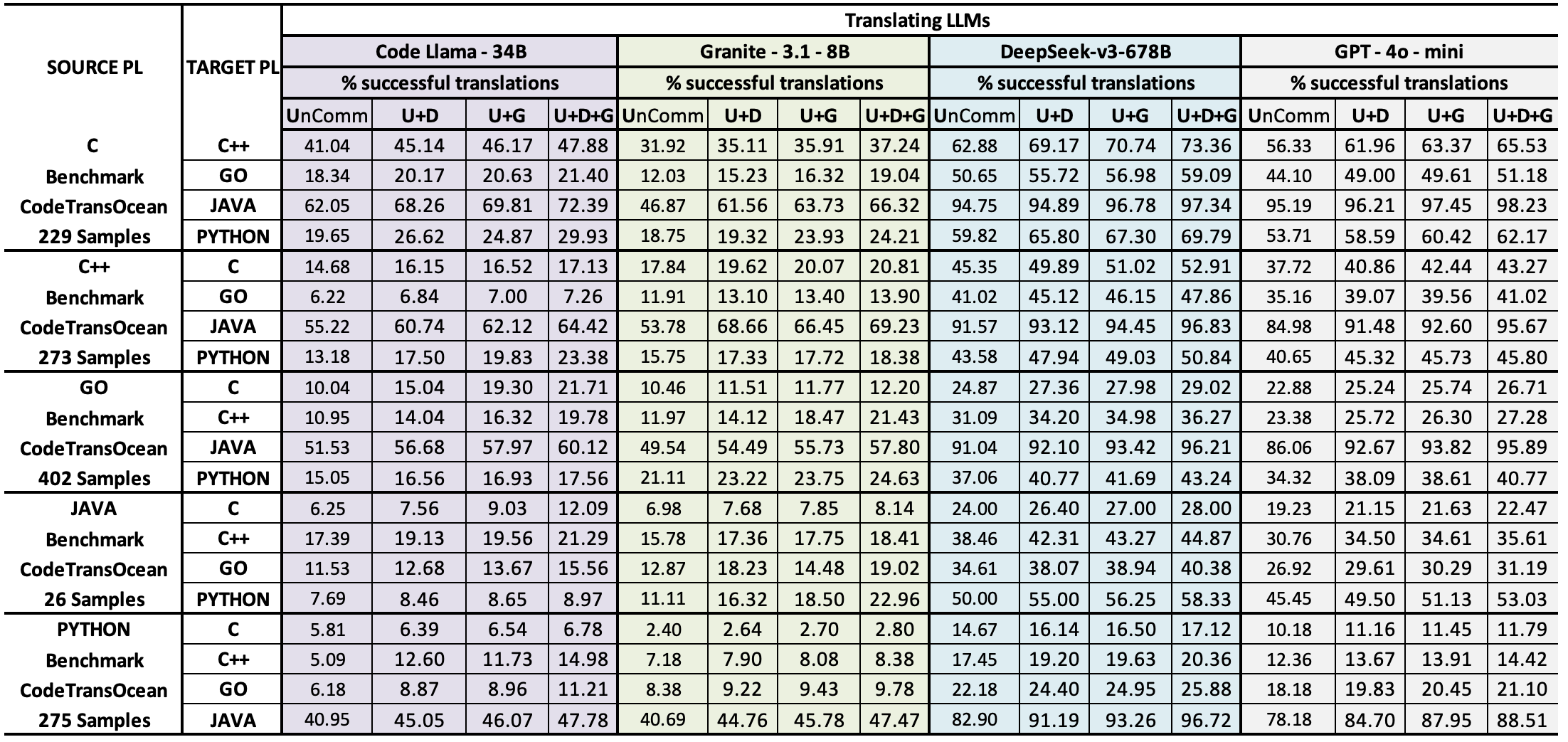}
  \caption{Code translation performance of various LLMs on all CodeTransOcean code samples. Note that newer versions of translating and commenting LLMs (mentioned in the corresponding column heading) were used in these experiments.}
  \label{fig:codetrans}
\end{figure}

CodeTransOcean \cite{codetransocean} 
 is a large-scale comprehensive multilingual benchmark that supports a
large variety of programming languages for
code translation. To rigorously assess the effectiveness of our proposed framework, COMMENTRA, we performed additional experiments on this benchmark, with the results presented in the tables below. Table \ref{tab:codetrans} presents the details of code sample extracted from the CodeTransOcean benchmark. All samples in our selected 5 PLs - C, C++, Go, Java, Python - which compiled and executed well and additionally had test cases embedded in them were extracted. In comparison to CodeNet and AVATAR benchmarks used earlier, code samples from this benchmark were slightly bigger in size in terms of lines of code. As with CodeNet and AVATAR benchmarks, none of the code samples had any author-written comments.
\vspace{-10pt}
\begin{table}
  \centering
  \begin{tabular}{cccc}
    \hline
    \multicolumn{4}{c}{\textbf{Dataset Statistics}} \\
    \hline
    Source PL & No. of  Samples & Min-Max LOC in samples\\
   \hline
    C & 229 & 3 - 199\\
    C++ & 273 & 2 - 250\\
    Go & 402 & 2 - 349\\
    Java & 26 & 5 - 194\\
    Python & 275 & 1 - 135\\
    \hline
    \multicolumn{4}{l}{Total Unique, Uncommented Code Samples : 1205} \\
    \hline
  \end{tabular}
  \caption{CodeTransOcean Benchmark Data Statistics}
  \vspace{-10pt}
  \label{tab:codetrans}
\end{table}
\vspace{-10pt}
Figure \ref{fig:codetrans} presents the code translation performance numbers of various translating LLMs on the code samples from CodeTransOcean benchmark. We observed that, for all translating models, the baseline uncommented code translation performance on these code samples was comparatively lower than that on the code samples from the AVATAR and CodeNet benchmarks. However, in tune with our observations on the AVATAR and CodeNet code samples, here also, repetitive translation iterations with various commenting LLMs fetched significant gains over the baseline uncommented translations.

\clearpage
\section{COMMENTRA - A Visualization of Performance Benefits due to GPT Comments on AVATAR and CodeNet Benchmarks, across all five PLs, for all five Translating Models}
\begin{figure}
\centering
\fbox{\includegraphics[scale=0.3]{ 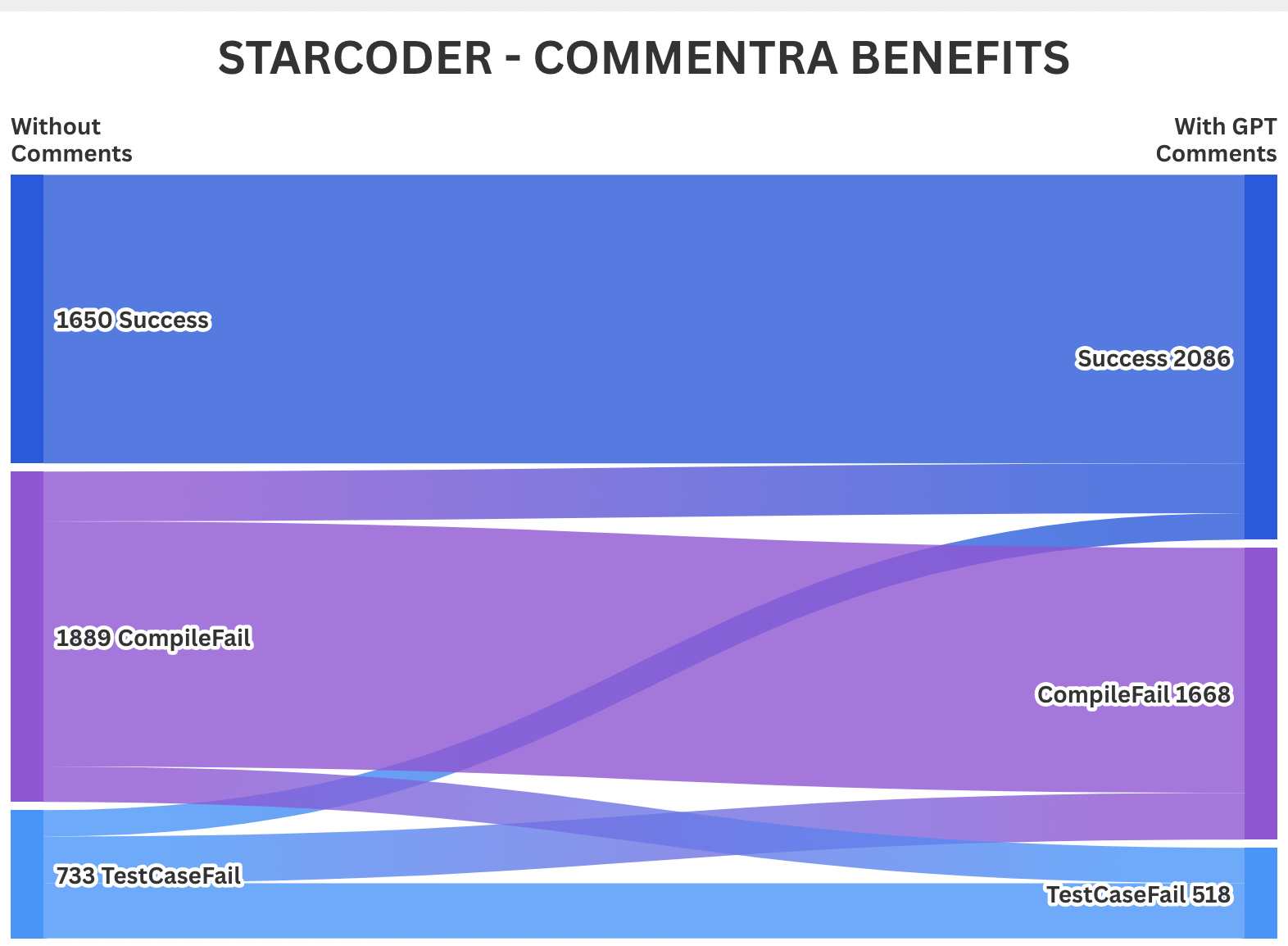}}
\end{figure}
\vspace{-40pt}
\begin{figure}
\centering
\fbox{\includegraphics[scale=0.3]{ 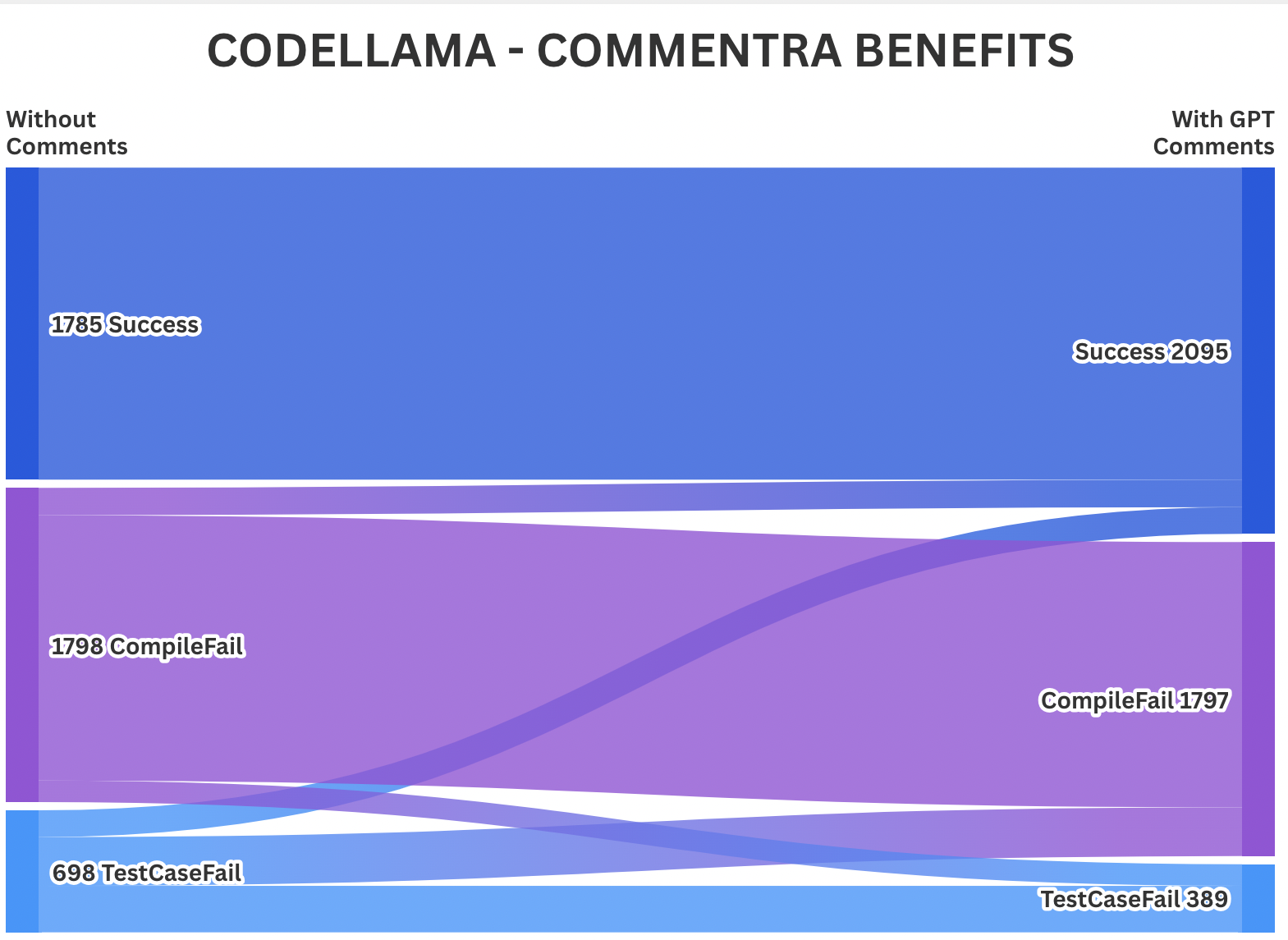}}
\end{figure}
\vspace{-40pt}
\begin{figure}
\centering
\fbox{\includegraphics[scale=0.28]{ 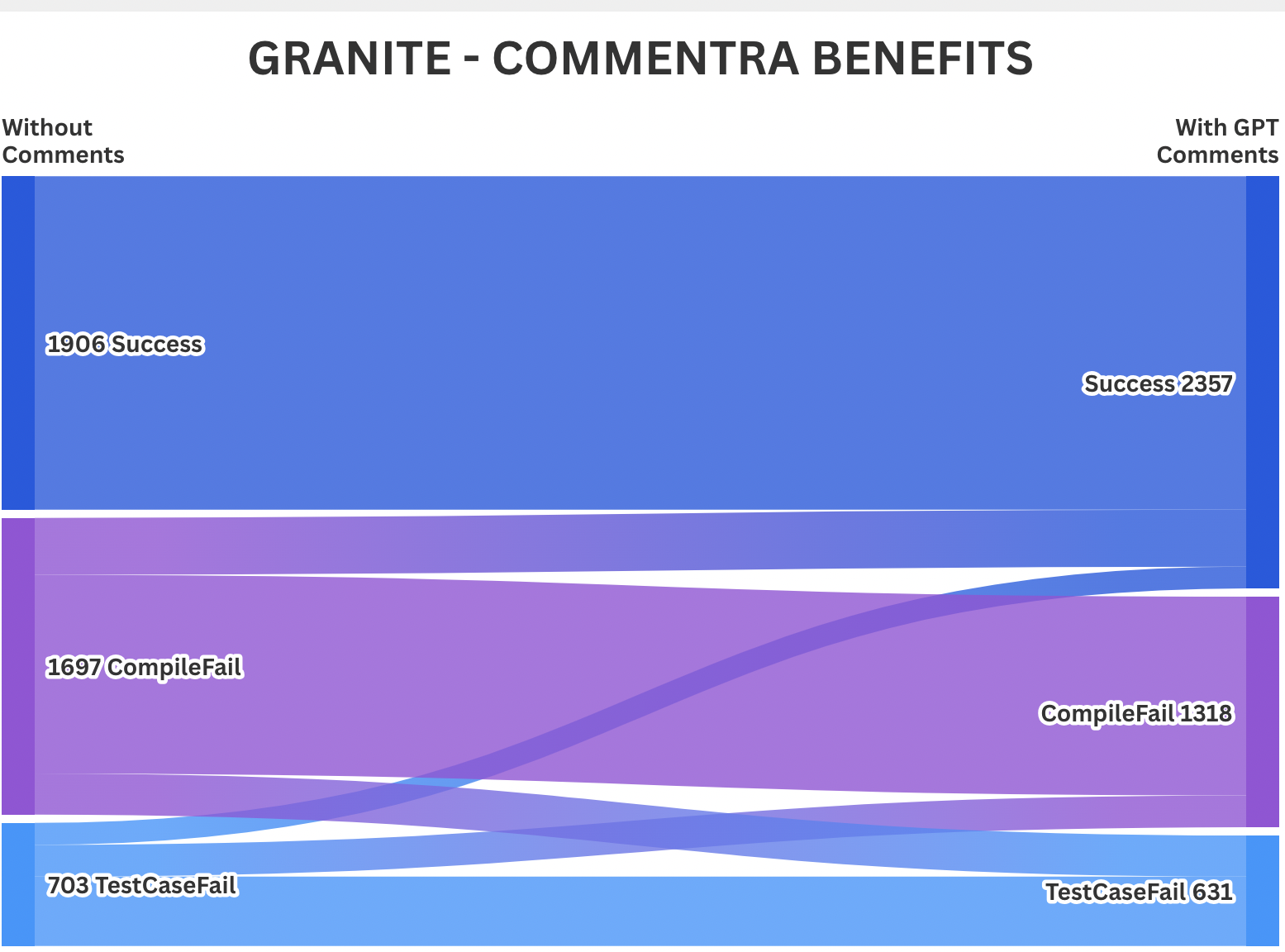}}
\end{figure}
\vspace{-40pt}
\begin{figure}
\centering
\fbox{\includegraphics[scale=0.28]{ 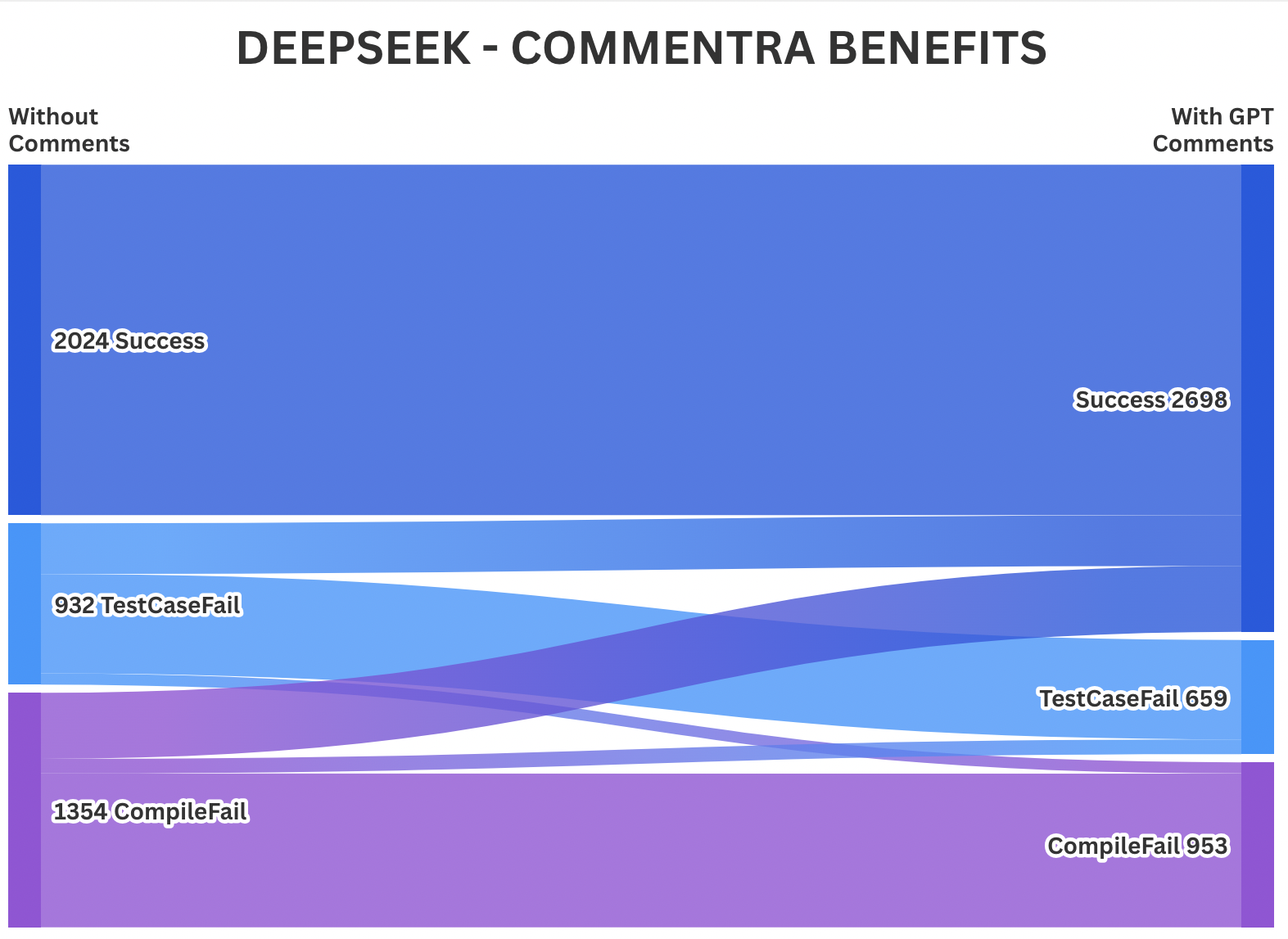}}
\end{figure}
\vspace{-40pt}
\begin{figure}
\centering
\fbox{\includegraphics[scale=0.28]{ 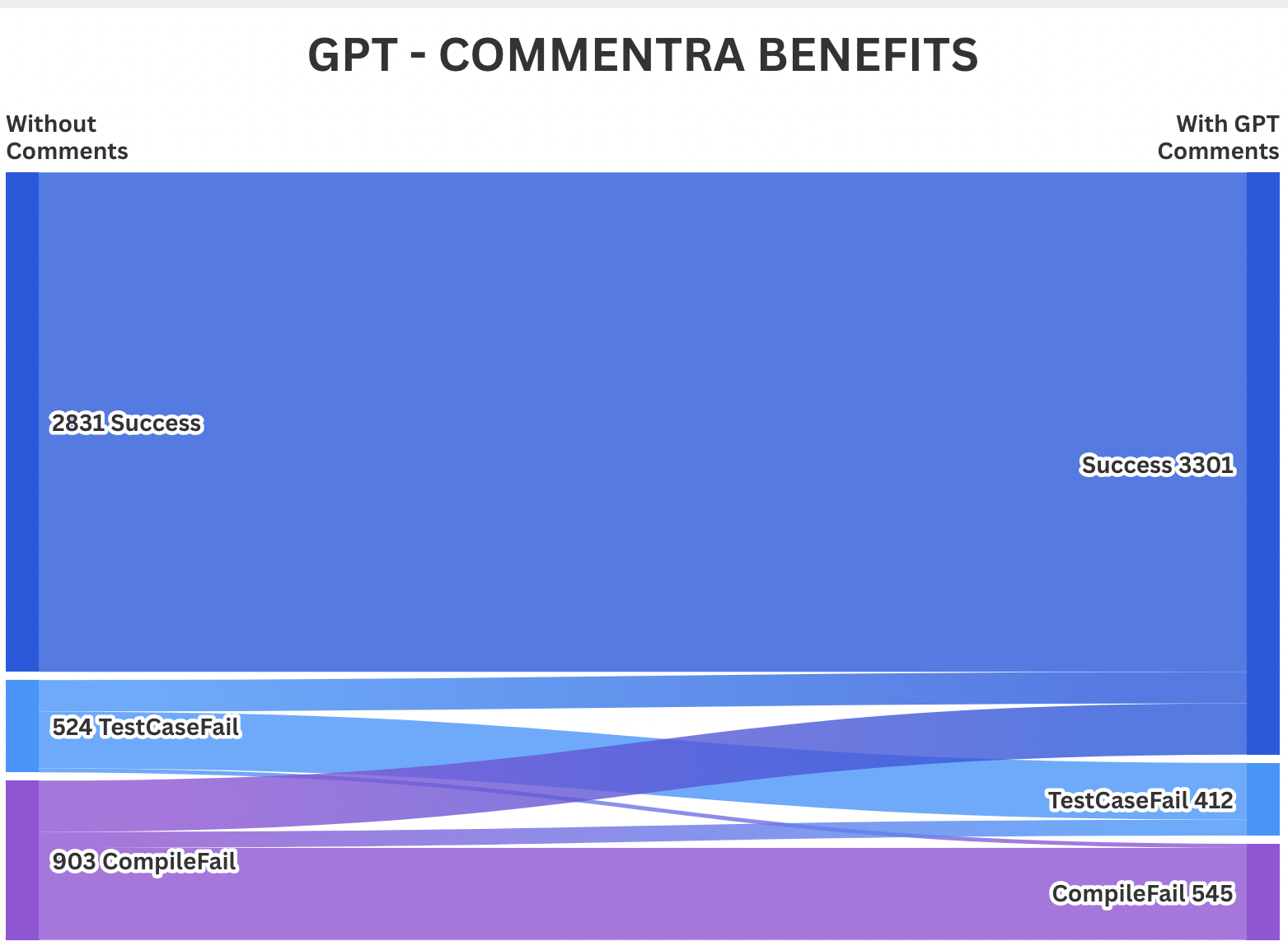}}
\end{figure}

\clearpage
\bibliographystyle{splncs04}
\bibliography{references}

\end{document}